\documentclass[sigconf]{acmart} %%camera ready

\AtBeginDocument{%
  }

\copyrightyear{2026}
\acmYear{2026}
\setcopyright{cc}
\setcctype{by}
\acmConference[CHI '26]{Proceedings of the 2026 CHI Conference on Human Factors in Computing Systems}{April 13--17, 2026}{Barcelona, Spain}
\acmBooktitle{Proceedings of the 2026 CHI Conference on Human Factors in Computing Systems (CHI '26), April 13--17, 2026, Barcelona, Spain}
\acmPrice{}
\acmDOI{10.1145/3772318.3790743}
\acmISBN{979-8-4007-2278-3/2026/04}

\definecolor{TODOcolor}{HTML}{FF0000}

\definecolor{inputvariable}{HTML}{1b9e77}
\newcommand{\inputvariable}[1]{\textcolor{inputvariable}{#1}}

\usepackage{soul}
\usepackage[normalem]{ulem}

\begin{document}

\title{The Impact of Uncertainty Visualization on Trust in Thematic Maps}

\author{Varun Srivastava}
\affiliation{%
  \institution{Arizona State University}
  \city{Tempe}
  \state{Arizona}
  \country{USA}
}
\email{vsriva11@asu.edu}
\orcid{0009-0002-7736-3574}

\author{Fan Lei}
\affiliation{%
  \institution{University of Waterloo}
  \city{Waterloo}
  \state{Ontario}
  \country{Canada}}
\email{fan.lei@uwaterloo.ca}

% \author{Michelle V. Mancenido}
% \affiliation{%
%   \institution{Arizona State University}
%   \city{Tempe}
%   \state{Arizona}
% \email{mvmancenido@asu.edu}

\author{Alan M. MacEachren}
\affiliation{%
 \institution{Pennsylvania State University}
 \city{University Park}
 \state{Pennsylvania}
 \country{USA}}
\email{maceachren@psu.edu}

\author{Ross Maciejewski}
\affiliation{%
  \institution{Arizona State University}
  \city{Tempe}
  \state{Arizona}
  \country{USA}}
\email{rmacieje@asu.edu}

\renewcommand{\shortauthors}{Srivastava et al.}

\begin{abstract}
Thematic maps are widely used to communicate spatial patterns to non-expert audiences. Although uncertainty is inherent in thematic map data, it is rarely visualized, raising questions about how its inclusion affects trust. Prior work offers mixed perspectives: some argue that uncertainty fosters trust through transparency, while others suggest it may reduce trust by introducing confusion. Yet few empirical studies explicitly measure trust in thematic maps. We conducted a between-subjects experiment (N = 161) to evaluate how visualizing uncertainty at varying levels (low, medium, high) influences trust. We find that uncertainty visualization generally reduces trust, with greater reductions observed as uncertainty levels increase. However, maps dominated by low uncertainty do not significantly differ in trust from those with no uncertainty. Moreover, while uncertainty visualization tends to make readers question the accuracy of the data, it appears to have a weaker influence on perceptions of the mapmaker’s integrity.
\end{abstract}





\maketitle

\section{Introduction}
\label{section1:introduction}

Thematic maps, such as choropleth maps and graduated symbol maps, are widely used to communicate spatial patterns in socio-economic, health, and environmental data~\cite{dent1999cartography, raposo2020change, juergens2020trustworthy}. These visualizations simplify complex geographic information and are increasingly consumed by non-expert audiences through journalism, public reports, and social media~\cite{roth2021cartographic, song2022visual, mocnik2020epidemics}. With the rise of accessible mapmaking tools, thematic maps are now created by a broader range of users—including data journalists, advocacy groups, and individuals without formal cartographic training~\cite{prestby2023trust, he2011visualize, griffin2020trustworthy}. While this democratization has expanded the reach of data-driven storytelling, it has also led to an influx of misleading or poorly designed visualizations, raising public skepticism toward the credibility of such maps~\cite{prestby2023trust, calfano2022bad}. In response, cartographers and visualization researchers have begun investigating how specific aspects of map design influence reader trust—among which uncertainty visualization has emerged as one of the key aspects~\cite{griffin2020trustworthy, prestby2023trust}.

In this paper, we use “uncertainty” to refer specifically to attribute uncertainty—variation in the quality or stability of underlying data values arising from sampling error, estimation noise, or reporting variability~\cite{thomson2005typology}. Although such uncertainty is inherent to spatial data~\cite{maceachren2005visualizing}, its impact on reader trust remains a topic of debate. Some scholars argue that visualizing uncertainty can enhance trust by promoting transparency and signaling honesty in representation~\cite{savelli2013advantages, padilla2022multiple}. Others caution that it may backfire—undermining user confidence, reducing data credibility, or overwhelming viewers with cognitive complexity~\cite{hullman2019authors}. In the domain of geovisualization, most prior research has focused on domain-specific applications—particularly in areas like meteorology and climate science—where the emphasis has been on improving decision-making or comprehension, often with expert users as the target audience~\cite{kinkeldey2017evaluating, kinkeldey2014assess}. Consequently, the impact of uncertainty visualization on non-expert readers' trust in thematic maps—where readers often lack the domain knowledge or statistical training needed to critically assess what they see—remains an open and underexplored topic.

To address this gap, our study empirically examines two key research questions: (1) Are thematic maps that visualize uncertainty more trustworthy than those that do not? (2) How does the level of uncertainty visualized in thematic maps affect trust? To our knowledge, this study represents the first empirical investigation to systematically measure how trust in thematic maps is modulated when uncertainty is visualized.

We build on the MAPTRUST framework proposed by Prestby~\cite{prestby2024measuring}, which offers an empirical approach to measuring trust in maps. This framework conceptualizes trust as a multidimensional construct comprising accuracy, dependability, and reliability, and operationalizes these dimensions through a set of 12 carefully selected adjectives. Using this framework, we designed a between-subjects experiment in which one group of participants viewed maps without uncertainty visualization, while another group saw maps that visualized uncertainty using symbol fuzziness at three ordinal levels (low, medium, high). The maps spanned six commonly used thematic topics: social, environment, health, crime, economic, and housing~\cite{dent1999cartography}. After viewing each map, participants rated it across the MAPTRUST adjectives, allowing us to quantitatively assess variations in trust across map conditions.

Our empirical findings reveal several key insights:
\begin{itemize}
    \item Uncertainty visualization has a measurable negative effect on trust in thematic maps, and this effect intensifies with increasing levels of uncertainty.
    \item Low levels of visualized uncertainty appear trust-neutral rather than trust-enhancing—they do not significantly boost trust but also do not diminish it relative to maps without uncertainty.
    \item The impact of uncertainty visualization appears to concentrate more on accuracy-related perceptions than on those related to dependability or reliability. This suggests that while uncertainty cues may prompt skepticism about the data itself, they may not necessarily erode perceived integrity or trust in the map's author.
\end{itemize}

These findings suggest that uncertainty visualization in thematic maps should be applied judiciously, with consideration for the author's objectives and the level of uncertainty present. While we provide empirical insights into the trust implications of such design choices, the ultimate decision rests with practitioners who must weigh these considerations against their specific communication goals.

\section{Related Work}
\label{section2:related_work}

To situate our study, we review three strands of relevant literature. First, we examine how trust in maps has been defined and measured in prior work. Second, we survey empirical studies that consider uncertainty visualization as a medium for influencing trust, focusing on cartographic and geovisualization research to maintain conceptual clarity. Finally, we review contributions that conceptualize geospatial uncertainty and propose techniques for representing it effectively.

\subsection{Trust in Maps: Definitions and Measurement}
\label{subsection2.1:trust_in_maps}

Trust has emerged as a critical yet inconsistently defined construct in cartography and visualization research. While maps have long been regarded as authoritative sources of information~\cite{griffin2020trustworthy, propen2007visual}, the growing role of user-generated mapping and the proliferation of misinformation have heightened the need to understand what it means to “trust” a map~\cite{prestby2025evaluating}. Prior work has defined trust variously as subjective certainty in the correctness of a map~\cite{schiewe2013vertrauen}, belief in the mapmaker’s intentions~\cite{mcgranaghan1999web}, or willingness to rely on a map in decision-making contexts~\cite{skarlatidou2011understanding}. However, these definitions are often fragmented and lack consensus~\cite{prestby2023trust, griffin2020trustworthy}.

Recent advances in visualization research—most notably Elhamdadi et al.'s Vistrust framework~\cite{elhamdadi2023vistrust}—emphasize that trust is inherently multidimensional, involving both cognitive assessments (e.g., perceived accuracy, clarity, or usability) and affective responses (e.g., emotional reactions, perceived benevolence). Their work extends longstanding psychological theories that distinguish cognition-based from affect-based trust. It shows that visualization trust emerges from factors tied to the visualization itself, the underlying data, and reader-specific characteristics. It further demonstrates that trust is not merely about accuracy or ethical intent but also about how a visualization shapes ongoing beliefs—such as by influencing Bayesian prior updating or altering one's willingness to act on the information presented.

Complementing this work, research in HCI has shown that trust in visualizations is deeply shaped by the viewer's relationship to the data. Peck et al.~\cite{peck2019data} demonstrated that non-experts often judge trustworthiness based not only on visual features but also on how personally relevant or sensitive the data feels. Their findings highlight that individuals may trust—or distrust—a visualization depending on whether its topic aligns with lived experience, local knowledge, or personal concerns. Recent work further shows that viewers also infer trust from stylistic and contextual cues in visualization design, using visual tone and framing to judge both data credibility and author intent~\cite{fox2025quantifying, morgenstern2025visualization}. This emphasizes the importance of studying trust in thematic maps among general audiences, whose prior experiences and familiarity with topics may strongly modulate their trust responses.

To address these conceptual gaps, Prestby synthesizes trust research in cartography to clarify how trust in maps should be defined and measured~\cite{prestby2023trust, prestby2024measuring, prestby2025evaluating}. His work reinforces the distinction between \textit{credibility}—momentary evaluations of believability—and \textit{trust}, which is forward-looking and reflects a willingness to depend on a map based on expectations of its accuracy and the integrity of its production. Aligning with this perspective, Prestby defines trust in maps as \textit{"the willingness to depend on a map based on the expectation that it is accurate and that acting on its visualized information will not be detrimental"}~\cite{prestby2024measuring}. We adopt this definition in our study.

A wide range of empirical approaches have been used to measure trust, from single-item Likert ratings~\cite{skarlatidou2013guidelines, xiong2019examining} to behavioral proxies such as whether participants would act on the map's information~\cite{joslyn2012uncertainty}. Yet, as both Prestby and Elhamdadi et al. note, such measures often fail to capture the full multidimensional nature of trust. Addressing these limitations, Prestby's MAPTRUST scale~\cite{prestby2024measuring} proposes a validated, multi-item instrument grounded in theory and data. It captures twelve adjectives spanning two core psychological dimensions of trust: an \textbf{Accuracy} dimension (e.g., Accurate, Correct, Error-free, Authentic, Objective) and a \textbf{Dependability and Reliability} dimension (e.g., Reliable, Trustworthy, Credible, Reputable, Balanced, Fair, Honest). These dimensions map closely onto the broader cognitive–affective components described in visualization research~\cite{elhamdadi2023vistrust}, providing a comprehensive evaluation tool that aligns with contemporary theories of how users form trust judgments from visual information. We adopt this MAPTRUST framework as our primary measurement approach in the current study.

Taken together, these conceptual and methodological developments highlight how trust must be measured in ways that are both rigorous and sensitive to the unique challenges of thematic maps, particularly for non-expert audiences. In these contexts, design choices can play a powerful role in modulating trust. In the following section, we examine one such design element—uncertainty visualization—and explore how it relates to trust in thematic maps.

%Such maps are increasingly consumed by readers who often lack the domain knowledge or statistical training needed to critically evaluate what they see~\cite{}.

\subsection{Uncertainty Visualization and Trust in Maps}
\label{subsection2.2:trust_and_uncertainty_visulization}

Uncertainty visualization refers to the set of techniques used to represent the limitations, variability, or confidence associated with data~\cite{maceachren2005visualizing}. Such visualizations communicate where data may be incomplete, error-prone, or inherently variable, providing audiences with a more transparent account of the information being presented~\cite{kinkeldey2014assess}. In the context of trust, uncertainty visualization carries a dual potential: it may enhance credibility by signaling honesty about data limitations, but it can also erode trust if readers interpret uncertainty as unreliability or find the representations overwhelming~\cite{hullman2019authors, griffin2020trustworthy}.

Empirical studies that examine how uncertainty visualization shapes trust in maps remain scarce. The few existing examples highlight both the promise and the pitfalls of incorporating uncertainty. For instance, Joslyn et al.~\cite{joslyn2012uncertainty} found that predictive interval forecasts in weather maps improved precautionary decision-making and, importantly, maintained trust compared to deterministic forecasts, but their operationalization of trust relied on whether participants said they would use the same forecast type again. Similarly, Kübler et al.~\cite{kubler2020against} showed in a multicriteria decision-making task that the presence and type of uncertainty visualization in hazard prediction maps shaped housing choices, with risk-prone individuals interpreting uncertainty differently from more cautious ones—again treating trust as a behavioral outcome rather than a measured construct. By contrast, Padilla et al.~\cite{padilla2022multiple} explicitly asked participants to rate their trust in COVID-19 forecast maps and demonstrated that design decisions, such as the number of models displayed or whether confidence intervals were included, significantly shaped reported trust: trust initially increased with more forecasts but plateaued and even declined when visual complexity grew. Taken together, these studies suggest that the impact of uncertainty visualization on trust in maps depends on the type of uncertainty depicted, its magnitude, and the characteristics of the audience~\cite{maceachren2015visual}. Yet empirical research examining this relationship remains limited, with most studies addressing trust only indirectly through behavioral proxies rather than explicit measurement.

Recent work in cartography has emphasized the need to move beyond such proxy measures and to quantitatively assess how uncertainty visualization modulates trust in readers~\cite{griffin2020trustworthy, prestby2025evaluating}. Prestby~\cite{prestby2023trust} calls for research that treats trust as a measurable outcome rather than an implicit byproduct of decision-making tasks—a need that motivates the research questions guiding this study. To address these questions, our experimental design draws directly on foundational principles of geospatial uncertainty, which we discuss in the following section.

\subsection{Geospatial Uncertainty}
\label{subsection2.3:geospatial_uncertainty}

In our study, we draw on longstanding research in cartography, geographic information systems (GIS), and geovisualization that has examined how uncertainty in spatial data can be conceptualized and communicated. Insights from this body of work inform both the classification of uncertainty types and the selection of visualization techniques that underpin the design of our experiment. In what follows, we review two key strands of this literature: foundational contributions that define types of geospatial uncertainty, and structured frameworks that classify visualization techniques for representing uncertainty in GIS.

\subsubsection{Uncertainty Types in Geospatial Data}
\label{subsubsection2.3.1:uncertainty_types}

Foundational contributions in GIS distinguish multiple dimensions of uncertainty that arise in spatial data. Thomson et al.~\cite{thomson2005typology} proposed a widely used typology that classifies uncertainty across three dimensions of geographic information—attribute, spatial, and temporal—each of which can contain inaccuracies related to accuracy, precision, completeness, consistency, lineage, currency, credibility, subjectivity, or interrelatedness. Of these, attribute accuracy has received significant attention because it directly affects the reliability of maps used in public communication and policy contexts. For example, uncertain attribute values can meaningfully alter decision making in domains such as disaster management~\cite{seipel2017color}, healthcare~\cite{maceachren1995mapping}, and public safety~\cite{huang2019exploring}. Identifying the specific type of uncertainty at play is therefore essential for selecting an appropriate visualization strategy.

\subsubsection{Uncertainty Visualization Techniques in GIS}
\label{subsubsection2.3.2:uncertainty_techniques}

A substantial body of work has proposed structured frameworks for visualizing uncertainty in GIS. Kinkeldey et al.~\cite{kinkeldey2014assess} classified geospatial uncertainty visualizations along three dichotomies: coincident vs.~adjacent displays (whether uncertainty is shown on the same map or a separate one), intrinsic vs.~extrinsic encodings (whether uncertainty modifies the main representation or appears as additional elements), and static vs.~dynamic approaches (whether the representation is fixed or interactive). Coincident, intrinsic, and static representations remain the most commonly applied in practice due to their simplicity and low cognitive overhead~\cite{kinkeldey2017evaluating}. Complementary work by MacEachren et al.~\cite{maceachren2012visual} evaluated a range of symbol sets for representing different uncertainty categories, showing that abstract cues—particularly fuzziness—are often the most intuitive for conveying generalized uncertainty.

More recent visualization research has emphasized that uncertainty representations can actively influence how readers interpret the underlying data, not merely communicate limitations. Correll et al.~\cite{correll2018value} introduced \textit{value-suppressing uncertainty palettes}, which intentionally reduce visual emphasis on uncertain values to mitigate overinterpretation. Their findings highlight that uncertainty encodings can shift attention and reshape takeaways in nontrivial ways, especially in choropleth-style thematic maps. Extending this line of work, Ndlovu et al.~\cite{ndlovu2023taken} empirically examined how Bayesian surprise and uncertainty suppression influence map-based interpretations. They showed that different uncertainty strategies can meaningfully alter the messages viewers extract from a map, demonstrating that uncertainty visualization can shape cognitive processing itself—not just awareness of data quality.

These studies highlight that uncertainty visualization is not a neutral design choice: it can reshape which spatial patterns appear salient, influence perceived signal strength, and meaningfully alter the takeaways readers form from thematic maps. We leveraged these prior findings and insights to make design choices for the stimuli used in our experiment, aimed at maximizing the intuitive understanding of uncertainty and minimizing the cognitive load that uncertainty visualizations introduce.

\section{Study Design}
\label{sec3:study_design}

This section outlines the design of our empirical study. We begin by presenting our research questions and corresponding hypotheses (Section~\ref{subsec3.1:research_questions}). Next, we describe the key factors related to uncertainty visualization used in the study (Section~\ref{subsec3.2:uncertainty_factors}). We then detail the map stimuli and thematic variations developed in close collaboration with a cartographer (Section~\ref{subsection3.3:map_stimuli}). Finally, we describe the participant sample (Section~\ref{subsection3.4:participants}) and the survey procedure, including the trust measurement instrument used to assess participants' responses to the maps (Section~\ref{subsection3.5:survey_measures}).

\subsection{Research Questions}
\label{subsec3.1:research_questions}

While prior studies in geospatial uncertainty visualization have primarily treated trust as an implicit outcome—inferring it from user behavior such as decision-making—our study treats trust as an explicit construct, measured directly through participants’ responses using the MAPTRUST framework~\cite{prestby2023trust, prestby2024measuring}. We focus on thematic maps consumed by non-expert audiences, who may be especially susceptible to misinterpretation or misinformation~\cite{griffin2020trustworthy}. Specifically, our experiment investigates how visualizing uncertainty—and varying its level—affects trust in such maps. The research questions and hypotheses outlined below reflect this goal.

\noindent \textbf{Research Question 1 (RQ1):} Are thematic maps that visualize uncertainty more trustworthy than those that do not? 
\begin{itemize} 
\item \textbf{Hypothesis 1 (H1):} Visualizing uncertainty will reduce trust in thematic maps. 

--\textit{Explanation:} Visualizing uncertainty may negatively impact trust due to increased cognitive load, confusion, and/or perceived lack of clarity. In addition, explicitly showing uncertainty can make readers more aware that the data represented is not fully certain or definitive.
\end{itemize}

\noindent \textbf{Research Question 2 (RQ2):} How does the level of uncertainty visualized in thematic maps affect trust?

\begin{itemize} 
\item \textbf{Hypothesis 2a (H2a):} Trust in thematic maps will decrease as the level of uncertainty increases.

--\textit{Explanation:} As uncertainty becomes more visually prominent, it may increasingly signal to readers that the map is merely a representation of the data—not an objective truth. This heightened awareness, coupled with more visible ambiguity, can erode confidence in the map’s reliability or the precision of its content.

\item \textbf{Hypothesis 2b (H2b):}  Maps with a low level of uncertainty may receive higher trust ratings than those with no uncertainty visualization.

--\textit{Explanation:} Low levels of uncertainty may strike a balance—providing just enough transparency to signal honesty and data integrity, without significantly undermining perceptions of trustworthiness. This subtle inclusion may lead readers to trust the map more than one that omits uncertainty altogether, which could appear overly confident.

\end{itemize}

\subsection{Uncertainty Factors}
\label{subsec3.2:uncertainty_factors}
% a. Describe how uncertainty is operationalized in the study.
% E.g., fuzziness (visual noise) as the uncertainty representation method.
% b. Define the three uncertainty levels (Low, Medium, High).

\subsubsection{Uncertainty Type and Visualization Technique}
\label{subsubsection3.2.1:uncertainty_type_technique}

To design the stimuli used to test our hypotheses, we selected a single uncertainty type and a single visualization technique, guided by the literature reviewed in Section \ref{subsection2.3:geospatial_uncertainty}. Because empirical work explicitly measuring trust in thematic maps is nascent, this study intentionally prioritized internal validity over breadth by using a tightly controlled design in which these choices were held constant across all maps. For the \textbf{uncertainty type}, we used sampling error—an attribute uncertainty common in survey-based socio-economic data~\cite{spielman2014patterns, schofield2006survey}. For the \textbf{visualization technique}, we adopted the \textit{intrinsic}, \textit{coincident}, and \textit{static} representation and encoded uncertainty using \textit{fuzziness}, following prior work identifying it as an intuitive and general-purpose visual cue~\cite{maceachren2012visual}.

\subsubsection{Uncertainty Levels}
\label{subsubsection3.2.2:uncertainty_levels}

To test our hypotheses, we introduced a controlled manipulation of uncertainty levels in the map stimuli. At the level of individual states (i.e., the geographic units displayed), each state was assigned one of three ordinal uncertainty levels:

\begin{itemize}
    \item \textbf{Low Uncertainty:} Estimates based on larger or more consistent samples.
    \item \textbf{Moderate Uncertainty:} Estimates have some variability.
    \item \textbf{High Uncertainty:} Estimates based on smaller or more variable samples.
\end{itemize}

\begin{figure}[t]
\centering	
% 0.5 for double col
\includegraphics[width=0.5\linewidth]{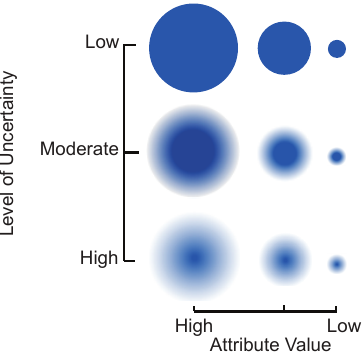}
\caption{Fuzziness design for uncertainty visualization. Each row corresponds to a different level of uncertainty (Low, Moderate, High), and each column represents an attribute value (High, Medium, Low). The circle fuzziness increases with uncertainty level while maintaining the underlying attribute value through size.}
\label{fig:fuziness}
\Description{This figure presents the uncertainty visualization design fuzziness in the form of graduated circles on maps. There are three levels: low, moderate, and high}
\end{figure}

Although numeric expressions of uncertainty (e.g., margins of error) are common in some domains, we intentionally used ordinal levels to maintain interpretability for non-expert audiences and isolate the perceptual effect of uncertainty magnitude. These levels enabled us to investigate not only whether uncertainty visualization affects trust, but also how the magnitude of uncertainty modulates this effect. The specific operationalization of these levels via geographic assignment is detailed in Section~\ref{subsubsection3.3.3:assigning_uncertainty_levels}. An example of how these uncertainty levels were visualized using fuzziness can be seen in Fig.~\ref{fig:fuziness}.

\subsection{Map Stimuli}
\label{subsection3.3:map_stimuli}

\begin{figure*}[t]
\centering	
% 0.9 for double col
\includegraphics[width=0.9\linewidth]{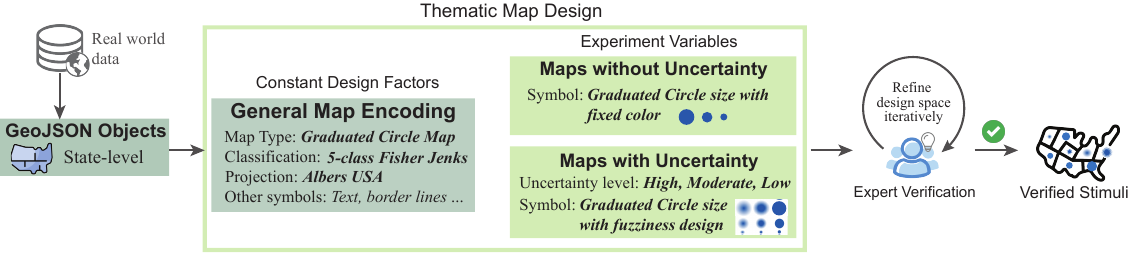}
\caption{Overview of the map stimuli design process. We used state-level real-world data in GeoJSON format to create thematic maps with consistent design factors (e.g., map type, classification, projection, and labeling). Two experimental conditions were defined: maps without uncertainty used fixed-color graduated circles, while maps with uncertainty used a fuzziness-based visual encoding with three levels (Low, Moderate, High). All maps were reviewed by a cartographer, and the final design space was refined iteratively to produce verified stimuli.}
\label{fig:stimuli_pipeline}
\Description{The figure illustrates the workflow used to generate the map stimuli used in the experiment. We began by sourcing real-world state-level data in GeoJSON format across six thematic topics. The map encoding followed a standardized set of design factors: all maps were rendered as graduated circle maps using a five-class Fisher Jenks classification and projected using the Albers USA projection. Constant design elements included textual labels, boundary lines, and other common map symbols. The experiment manipulated one key design variable: uncertainty visualization. In the “Maps without Uncertainty” condition, circle size was used to represent data magnitude, and a fixed color was applied. In the “Maps with Uncertainty” condition, the same symbol size encoding was retained, but visual fuzziness was added to communicate uncertainty. Each map with uncertainty represented one of three dominant uncertainty levels—High, Moderate, or Low—determined based on the proportion of states assigned to each level. The entire design space was iteratively refined in collaboration with a cartographic expert. Finalized map stimuli were verified before inclusion in the experimental survey.}
\end{figure*}

This section describes the process used to generate the map stimuli shown to participants. We began by aggregating real-world state-level data in GeoJSON format (Section~\ref{subsubsection3.3.1:map_theme_selection}) and encoded it using a standardized set of design parameters. We then defined the base map design (Section~\ref{subsubsection3.3.2:base_map_design}), and finally encoded uncertainty levels (High, Moderate, Low) using a fuzziness-based design (Section~\ref{subsubsection3.3.3:assigning_uncertainty_levels}). An illustrative overview of this pipeline is provided in Fig.~\ref{fig:stimuli_pipeline}.

\subsubsection{Map Theme Selection} 
\label{subsubsection3.3.1:map_theme_selection}

The thematic content of a map is relevant in the context of our study, as uncertainty cues can modulate readers’ reliance on prior knowledge about the visualization topic during interpretation~\cite{hullman2019authors, karduni2020bayesian}. To account for such variation and prevent confounding effects, we treated \textbf{map theme} as a controlled variable in our experimental design. We selected six themes that are widely studied and commonly represented in thematic cartography~\cite{dent1999cartography, slocum2022thematic}: social, economic, health, crime, environment, and housing.

\begin{figure}[t]
\centering	
% just \linewidth for double col
\includegraphics[width=\linewidth ]{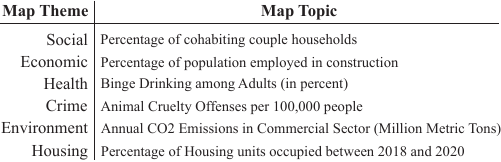}
\caption{The six thematic categories used in the study, along with their corresponding map topics. Data for Social, Economic, and Housing themes were sourced from the U.S. Census Bureau~\cite{us_census_acs}; Health data from the CDC~\cite{cdc_data}; Crime data from the FBI Crime Explorer~\cite{fbi_crime_data_explorer}; and Environment data from the EPA~\cite{epa_data}. All datasets were from the year 2024.}
\label{fig:table_map_theme}
\Description{This table outlines the six thematic categories (map themes) used in the experiment, each paired with its specific map topic. The Social, Economic, and Housing datasets were sourced from the U.S. Census; Health data from the Centers for Disease Control and Prevention (CDC); Crime data from the FBI Crime Explorer; and Environmental data from the Environmental Protection Agency (EPA). All data were obtained from 2024 public datasets. These themes were selected to ensure a diverse representation of real-world socio-economic and environmental issues relevant to U.S. states.}
\end{figure}

One dataset was selected for each theme, using only data from the United States. The data were sourced from reputable organizations, including the U.S. Census Bureau~\cite{us_census_acs} and the Environmental Protection Agency (EPA)~\cite{epa_data}. Further details are provided in Fig.~\ref{fig:table_map_theme}.

\subsubsection{Base Map Design}
\label{subsubsection3.3.2:base_map_design}
To ensure consistent map design and minimize potential perception biases, we adhered to established thematic design principles throughout the experiment~\cite{10273434}. We classified the attribute values in each dataset into five distinct classes using the well-regarded Fisher-Jenks method~\cite{jenks1971error}. This method effectively addresses deviations around class means, facilitating a nuanced understanding of data distribution across maps, as demonstrated in prior work~\cite{fan2024understanding}. Aligned with our chosen uncertainty visualization technique (detailed in Section~\ref{subsubsection3.2.1:uncertainty_type_technique}), proportional symbols were employed to represent the classified attribute values. A blue color scheme was selected using Colorbrewer~\cite{brewer2016designing}. Legends displayed the attribute classes along with their corresponding value ranges, and included a note stating: “The data shown are based on samples.”

\begin{figure*}[t]
\centering	 
% 0.9 for double col
\includegraphics[width=0.9\linewidth]{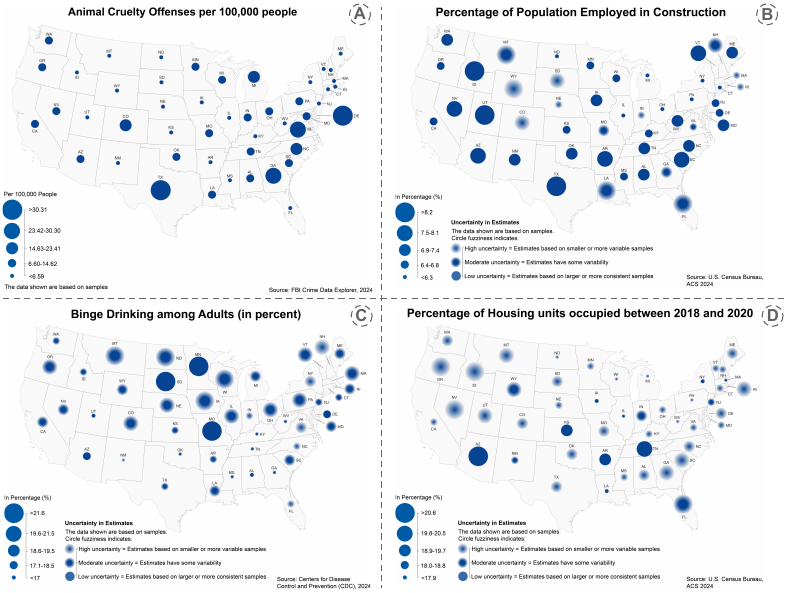}
\caption{Sample maps used in the experiment. Panel A shows a thematic map without uncertainty visualization, while Panels B, C, and D depict maps with uncertainty, with low (B), medium (C), and high (D) levels of uncertainty dominance, respectively.}
\label{fig:map_stimuli}
\Description{This figure presents four sample thematic maps used as stimuli in the experiment, each representing different uncertainty conditions. Panel A displays a baseline map without any uncertainty visualization and was shown to participants in Group A. Panels B, C, and D include visual representations of uncertainty and were shown to participants in Group B. The uncertainty was introduced using a fuzzy, semi-transparent overlay design. These overlays were applied selectively across states to reflect varying dominance levels of uncertainty: Panel B shows a map with low uncertainty dominance (approximately 70\% of states having low uncertainty), Panel C reflects medium dominance (about 70\% of states with medium uncertainty), and Panel D represents high uncertainty dominance (with high uncertainty present in the majority of states). All maps used the same base layout and symbology, and only the uncertainty overlays differed across panels. This controlled design allowed for isolating the effect of uncertainty level on participants' trust perceptions while holding other map features constant.}
\end{figure*}

To reduce potential trust biases associated with data provenance~\cite{prestby2023trust}, each map also cited the data source clearly in the bottom-right area of the layout. Further enriching the maps for clear communication, we incorporated additional encodings following established design guidelines~\cite{dent1999cartography}. Standard two-letter state abbreviations were included in an 18-point Arial font, and state borders were drawn with a line width of 0.5 points. For states with limited space for text labels, lead lines of 1-point width were used to connect the abbreviation to the corresponding state location. Consistency was maintained across all maps through the use of the Albers USA projection~\cite{snyder1982map}, yielding six univariate proportional symbol maps referred to as \textbf{“Maps without Uncertainty”} (Fig.~\ref{fig:map_stimuli}A).

\subsubsection{Assigning Uncertainty Levels}
\label{subsubsection3.3.3:assigning_uncertainty_levels}

To test the role of uncertainty magnitude in shaping trust, we introduced the notion of \textbf{uncertainty dominance level}—a key design factor in our study. Whereas individual states are assigned an uncertainty level (Low, Medium, or High; see Section \ref{subsubsection3.2.2:uncertainty_levels}), a map’s uncertainty dominance refers to the overall distribution of those levels across the map as a whole. A map was considered dominated by a particular uncertainty level when that level was assigned to the majority of geographic units displayed. For example, a “high-dominant” map featured a distribution where most U.S. states exhibited high uncertainty. 

To operationalize this, we used synthetic uncertainty assignments rather than real-world sampling error to maintain experimental control and create clearly distinguishable uncertainty conditions. Real-world uncertainty values can exhibit strong spatial clustering—driven by population size, sampling design, and topic-specific data sparsity—which would introduce confounds unrelated to uncertainty magnitude. Using synthetic assignments allowed us to isolate the causal effect of uncertainty level on trust while avoiding unintended geographic or thematic bias.

Each map was constructed such that approximately 70\% of the states were assigned the uncertainty dominance level (\textbf{Low, Medium, or High}), with the remaining 30\% divided between the other two levels. To further avoid geographic bias, uncertainty assignments were balanced across U.S. regions based on the U.S. Census Bureau’s four-region schema~\cite{uscb_econ_geo_levels}: West, Midwest, South, and Northeast\footnote{We excluded Alaska, Hawaii, and Puerto Rico due to their non-contiguous locations, as well as the District of Columbia}. This regional balancing ensured that observed trust differences were not inadvertently driven by region-specific uncertainty patterns—for example, the systematically higher sampling error typical of small-population states.

State-to-uncertainty assignments were randomized within these constraints using Python (code provided in the Supplementary Materials). Appendix~\ref{appendixA1:region_groupings} lists the full region definitions.

After assigning uncertainty levels, the fuzziness-based uncertainty visualization (see Fig.~\ref{fig:fuziness}) was applied. Each map included a legend indicating the mapping between circle fuzziness and uncertainty level, with the definition of each level clearly stated (as described in Section~\ref{subsubsection3.2.2:uncertainty_levels}). This process yielded 18 \textbf{"Maps with Uncertainty"}—one for each combination of the 6 themes and 3 uncertainty dominance levels. Example maps are shown in Fig.~\ref{fig:map_stimuli} (B, C, D), while all the maps used in the experiment are provided in the Supplementary Materials.

\subsection{Participants}
\label{subsection3.4:participants}

Participants were recruited through Prolific~\cite{prolific2025}, restricted to U.S.-based, English-fluent individuals with at least a 95\% approval rating and five prior submissions. The survey took approximately 20 minutes, and participants received \$5 compensation (equivalent to \$15/hour). The study was approved by an Institutional Review Board (IRB) and complied with all ethical guidelines for research involving human subjects. After removing those who failed a basic attention check, 161 participants remained (86 female, 73 male, 2 nonbinary; mean = 41.9, std-dev = 11.7). All had at least a high school education: 55 held a GED or equivalent, 55 held a four-year degree, and the rest held postgraduate degrees.

\subsection{Survey Measures}
\label{subsection3.5:survey_measures}

\begin{figure*}[t]
\centering	 
% 0.9 for double col
\includegraphics[width=0.9\linewidth]{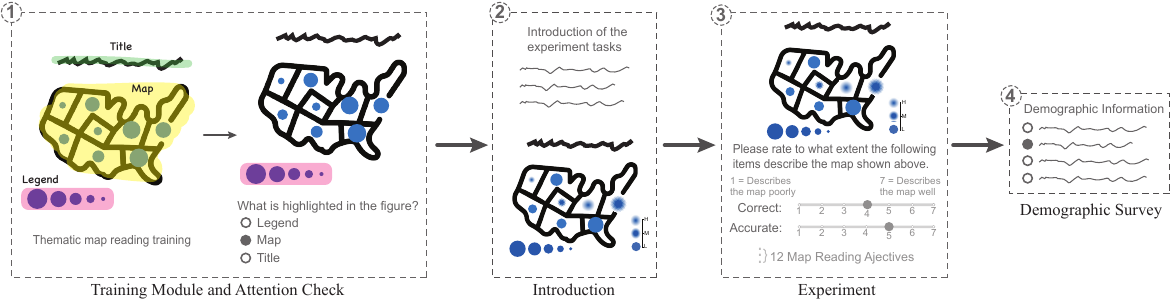}
\caption{The survey comprised four key stages: 1) Basic map elements training and attention check, 2) Introduction to main experiment, 3) Rating maps on the MAPTRUST scale, 4) Demographic Survey}
\label{fig:survey_flow}
\Description{The figure illustrates the flow of the user study. It consists of four sequential modules: (1) Training Module and Attention Check, where participants are introduced to map elements (title, legend, geographic area) and complete a visual attention check; (2) Introduction, which provides a brief overview of the task; (3) Experiment, where participants view maps and rate them using twelve MAPTRUST adjectives (e.g., Accurate, Honest, Credible); and (4) Demographic Survey, where participants provide background information such as age, gender, and geographic familiarity.}
\end{figure*}

As shown in Fig.~\ref{fig:survey_flow}, the full survey flow was designed to assess how participants evaluated thematic maps based on 12 MAPTRUST adjectives~\cite{prestby2024measuring}, which together capture dimensions of trust in maps (see Section~\ref{subsection2.1:trust_in_maps}). Before beginning the survey, participants were randomly assigned to one of two groups (A vs. B). Group A participants interacted solely with “Maps without Uncertainty,” while Group B participants engaged only with “Maps with Uncertainty.”

The survey began with a short training module to familiarize participants with basic map-reading components. A sample map was shown with labeled elements including the \textbf{title}, \textbf{legend}, and \textbf{base map}. Participants were then asked two attention-check questions requiring them to identify features from the training map. Those who answered either question incorrectly were disqualified from proceeding further.

Next, participants received brief instructions introducing the main experiment. They were told: \textit{“You will be shown six maps. For each map, please rate how well each descriptive word applies to the map overall—based on how it looks and the data it presents. There are no right or wrong answers; we are interested in your personal impressions of the map as a whole.”\footnote{We chose this wording to reduce judgment pressure and encourage honest, subjective evaluations—crucial when measuring trust-related impressions.}}

Participants then proceeded through six maps shown in randomized serial order. A split-plot experimental design~\cite{montgomery2017design} was implemented to ensure balanced exposure to map variations while controlling for order effects and survey fatigue. Block randomization was applied independently for each group.
\begin{itemize}
\item \textbf{Group A} participants viewed six “Maps without Uncertainty”—one from each of the six thematic topics.
\item \textbf{Group B} participants only saw maps with uncertainty—one from each theme. Crucially, their map set included two maps, each dominated by low, medium, and high uncertainty levels.
\end{itemize}

Each map shown to participants was followed by a question asking them to rate how well twelve descriptive adjectives applied to that map. These adjectives, drawn from the MAPTRUST instrument~\cite{prestby2024measuring}, were: \textit{Accurate}, \textit{Correct}, \textit{Error-free}, \textit{Honest}, \textit{Trustworthy}, \textit{Credible}, \textit{Fair}, \textit{Reliable}, \textit{Reputable}, \textit{Objective}, \textit{Authentic}, and \textit{Balanced}. Participants responded using a 7-point Likert scale, where 1 indicated “describes the map very poorly” and 7 indicated “describes the map very well.”  To mitigate order effects~\cite{krosnick1987evaluation}, the sequence in which the adjectives were presented was randomized for each participant. However, once this order was established, it remained consistent across all six maps viewed by that participant. 

\begin{figure}[t]
\centering	
% only \linewidth for double col
\includegraphics[width=\linewidth ]{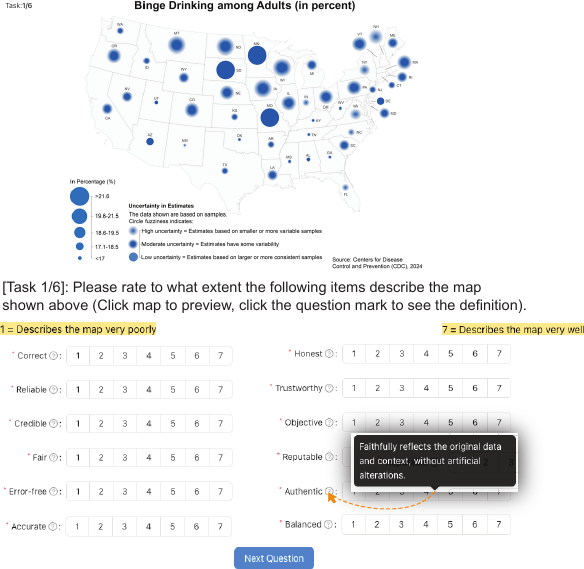}
\caption{Rating interface shown to participants during the main experiment. Each map was accompanied by twelve MAPTRUST adjective prompts, which participants rated on a 7-point Likert scale. A tooltip icon (“?”) was available next to each adjective to provide definitions as needed.}
\label{fig:rating_interface}
\Description{The figure shows the rating interface presented to participants during the main experiment. At the top of the interface is a thematic map titled “Binge Drinking among Adults (in percent),” with circular markers of varying sizes and colors overlaid on U.S. states to indicate attribute values. Below the map, twelve trust-related adjectives from the MAPTRUST instrument are listed in a 2-column grid format. Each adjective is followed by a 7-point Likert scale ranging from 1 ("Describes the map very poorly") to 7 ("Describes the map very well"). To the right of each adjective is a small question mark (“?”) icon that, when hovered over, provides a tooltip definition of the adjective. A “Next Question” button appears at the bottom of the interface to proceed to the next trial.}
\end{figure}

To minimize confusion between semantically similar adjectives (e.g., \textit{accurate}, \textit{correct}, and \textit{error-free}), participants were given the option to view brief definitions of each adjective by clicking a tooltip “?” icon next to the word. These definitions were adapted from prior literature on trust in visualizations; additional details and citation sources are provided in Appendix~\ref{appendixA2:maptrust_adjectives}, while an example of the rating interface is shown in Fig.~\ref{fig:rating_interface}. 
The survey concluded with a short demographic questionnaire. Participants were asked to report their age, gender, highest level of education completed, and self-reported familiarity with map reading.

\section{Results}

\subsection{Data and Models}
\label{subsection4.1:data_and_models}
% Purpose: Sets up the foundation by explaining:
% 1. How raw responses were processed and cleaned
% 2. How Likert responses were averaged/rounded (if applicable)
% 3. Justification for choosing the main modeling technique: Ordinal Logistic Regression, likely using a cumulative link mixed model (CLMM) 
% 4. Any software/libraries used ( ordinal package in R)
% 5. Mention any checks for assumptions (e.g., proportional odds assumption if relevant)
% 6. Clarify random effects used (e.g., participant as a random intercept)
% 7. Note how MAPTRUST adjective scores were aggregated (e.g., average per map, per participant)
\subsubsection{Data Preprocessing}
\label{subsubsection4.1.1:data_processing}

The final dataset comprised responses from 161 participants: \textbf{82 in Group A} (Maps without Uncertainty) and \textbf{79 in Group B} (Maps with Uncertainty), yielding a total of \textbf{966} map-level responses (6 maps × 161 participants). To prepare the data for statistical modeling, we computed a composite trust score by averaging the 12 individual MAPTRUST Likert ratings provided by the participants for each map. This average was then rounded to the nearest integer to generate a single ordinal measure, referred to as \textbf{trust\_overall}. This variable, ranging from 1 (lowest) to 7 (highest), served as the main \textbf{dependent variable} in our statistical models.

\subsubsection{Statistical Models}
\label{subsubsection4.1.2:statistical_models}

To model the ordinal \textit{trust\_overall} scores (ranging from 1 to 7), we employed a \textbf{Cumulative Link Mixed Model (CLMM)} with a logit link function~\cite{mccullagh1980regression, agresti2010analysis, greene2010modeling}, implemented via the ordinal package in R~\cite{christensen2018cumulative}. This modeling approach is appropriate for ordinal outcomes and allows for both fixed effects (i.e., predictor variables) and random effects to account for participant-level variability, as each participant contributed six observations.

At its core, the CLMM estimates the cumulative probability of a response falling at or below each threshold of the ordinal outcome. Formally, the model is defined as:
\begin{equation}
\label{equation:clmm}
\text{logit} \left( P(Y \leq k \mid X) \right) = \theta_k - \left( \mathbf{X\beta} + b_i \right), \quad \text{for } k = 1, \dots, 6
\end{equation}
Here, $Y$ is the ordinal response variable (e.g., trust\_overall); $k$ indexes the six cumulative thresholds between adjacent Likert scores; $\theta_k$ are threshold parameters to be estimated; $\mathbf{X}$ is the vector of fixed-effect predictors (e.g., group or uncertainty dominance); $\boldsymbol{\beta}$ are the corresponding fixed-effect coefficients representing the log-odds of being in a higher trust category; and $b_i$ is a random intercept capturing participant-level variation.

This framework allows us to interpret the exponentiated coefficients, $exp(\beta)$, as \textbf{odds ratios}. For example, an odds ratio less than 1 indicates lower odds of assigning a higher trust rating with increasing levels of the predictor, while a ratio greater than 1 indicates higher odds. All hypothesis tests were conducted at a \textbf{significance level of $\alpha = 0.05$.} Full model specifications and R code are provided in Supplementary Materials to support reproducibility. 

\subsection{Hypothesis Testing}
\label{subsection4.2:hypothesis_testing}
% Model coefficients, odds ratios, p-values

% Visuals: predicted probabilities or margins plots are very helpful

% Summary interpretation (e.g., “Participants were 2.3× less likely to rate maps with high uncertainty as trustworthy compared to those with low uncertainty.”)
\noindent \textbf{Hypothesis 1 (H1): Visualizing uncertainty will reduce trust in thematic maps.}

To test H1, we fit the CLMM described in Equation~\ref{equation:clmm}. The response variable ($Y$) was the ordinal trust score \textbf{trust\_overall}. The model included two fixed effects: \textbf{\inputvariable{Group}} and \textbf{\inputvariable{Map Theme}}. The former captured the primary effect of interest—whether the presence or absence of uncertainty visualization affects trust (Group A = maps without uncertainty vs. Group B = maps with uncertainty). The latter served as a control variable to account for potential theme-specific influences across the six map themes, ensuring that any observed differences could be attributed to uncertainty visualization irrespective of map content.

\begin{figure}[t]
\centering	
% only \linewidth for double col
\includegraphics[width=\linewidth ]{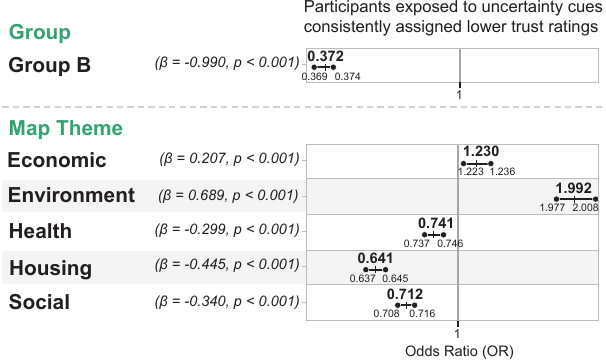} %table was hypothesis1.pdf
\caption{CLMM results for H1: Odds ratios (with 95\% confidence intervals) for the effect of Group and Map Theme on the odds of assigning a higher trust rating. Values are shown relative to the reference levels (Group A and Crime, not shown). The coefficient $\beta$ corresponds to the model’s estimated log-odds effect, and the plotted odds ratios represent exp($\beta$). The vertical reference line at OR = 1 indicates no difference from the reference level; odds ratios below 1 reflect lower likelihood of assigning higher trust.}
\label{fig:table_H1_results}
\Description{This figure displays the odds ratios and 95\% confidence intervals estimated from the CLMM testing H1, showing how the presence of uncertainty visualization (Group B) and the map’s thematic topic affect the odds of assigning a higher trust rating. Each horizontal interval bar conveys the magnitude and direction of the effect relative to the reference levels (Group A and the Crime theme). The visual layout separates the Group effect from the Theme effects, allowing readers to compare their relative strength. The plot highlights that the effect for Group B is substantially below 1.0, indicating lower trust when uncertainty is shown. The figure assists readers in visually interpreting effect sizes, confidence intervals, and the comparative importance of predictors in the model.}
\end{figure}

To interpret the model, we note that \textit{Group A} served as the reference category for Group, and the \textit{Crime} theme was the reference level for Map Theme\footnote{The reference levels for both variables were assigned by default based on alphabetical order by the R modeling function}. The results—shown in Fig.~\ref{fig:table_H1_results}—present the odds ratios with 95\% confidence intervals, along with the corresponding estimated coefficients for each predictor.

As shown in the figure, \textbf{Group B had a statistically significant negative effect on trust ratings}, with an estimated coefficient of –0.990, corresponding to an odds ratio of 0.372 (p < .001). This indicates that, regardless of map theme, participants who viewed maps with uncertainty were approximately 62.8\% less likely to assign higher trust ratings compared to those who viewed maps without uncertainty. Therefore, \textbf{we fail to reject H1.}

\vspace*{2pt}

\noindent \textbf{Hypothesis 2a (H2a): Trust in thematic maps will decrease as the level of uncertainty increases.}

To test H2a, we again used the CLMM framework described in Equation~\ref{equation:clmm}, with \textbf{trust\_overall} as the ordinal response variable. This model was fit exclusively on responses from \textbf{Group B participants}, as the goal was to examine how trust varies across different levels of uncertainty dominance. The model included two fixed effects: \textbf{\inputvariable{Uncertainty Dominance Level (UDL)}} and \textbf{\inputvariable{Map Theme}}, while other specifications remained the same as H1. In this model, \textit{Low} served as the reference level for UDL.

\begin{figure}[t]
\centering	
% only \linewidth for double col
\includegraphics[width=\linewidth ]{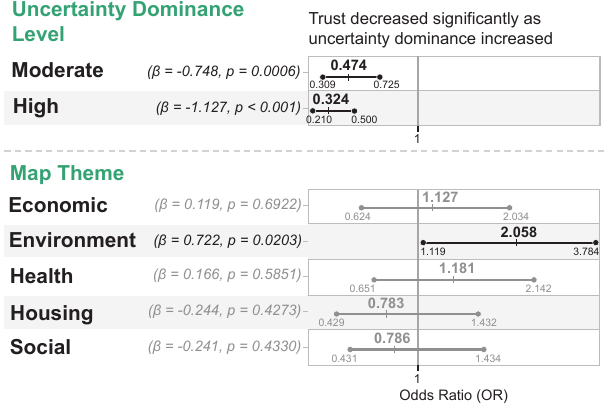}
\caption{CLMM results for H2a: Odds ratios (with 95\% confidence intervals) for the effect of Uncertainty Dominance Level (reference: Low, not shown) and Map Theme (reference: Crime, not shown) on the odds of assigning a higher trust rating. The corresponding coefficient estimates ($\beta$) and p-values are shown alongside each predictor. Statistically significant values ($\alpha$ = 0.05) are shown in \textbf{bold}.}
\label{fig:table_H2a_results}
\Description{This figure displays the CLMM results for Hypothesis 2a, showing how trust varies across uncertainty dominance levels within Group B. The top panel presents odds ratios for Moderate and High uncertainty dominance relative to the Low reference level. Both effects are visually separated and include 95\% confidence intervals, with statistically significant values highlighted in bold. The plotted intervals illustrate that Moderate and High uncertainty conditions shift the odds ratios well below 1, indicating reduced likelihood of assigning higher trust ratings as uncertainty increases. The lower panel displays the effects of the five non-reference map themes, each shown with its corresponding odds ratio and confidence interval. The x-axis is shared across both panels and encodes the odds-ratio scale, enabling direct comparison across predictors.}
\end{figure}

As shown in results from Fig.~\ref{fig:table_H2a_results}, \textbf{both Moderate and High UDLs had statistically significant negative effects on trust ratings}:
\begin{itemize}
    \item For maps dominated by moderate uncertainty, the estimated coefficient was –0.748 (odds ratio = 0.473, p = .0006), indicating participants were about 52.6\% less likely to assign higher trust ratings compared to maps with Low UDL.
    \item For maps dominated by high uncertainty, the coefficient was even more negative (–1.127, odds ratio = 0.324, p < .001), suggesting a 67.6\% reduction in odds of assigning higher trust ratings relative to Low UDL maps.
\end{itemize}

These findings indicate a clear and statistically significant decline in trust as the UDL increases, regardless of the thematic content of the map. Therefore, \textbf{we fail to reject H2a}.

\vspace*{2pt}

\noindent \textbf{Hypothesis 2b (H2b): Maps with low level of uncertainty
may receive higher trust ratings than those with no uncertainty visualization.}

To evaluate H2b, we returned to a between-group comparison similar to H1, with a specific focus on whether visualizing low uncertainty improves trust relative to omitting uncertainty altogether. This model was fit on responses from both Group A and Group B participants.

The response variable was again \textbf{trust\_overall}. Fixed effects included \textbf{\inputvariable{Map Theme}} and a four-level categorical predictor combining \textbf{\inputvariable{Group and UDL}}: Group A, Group B\_Low, Group B\_Moderate, and Group B\_High.

\begin{figure}[t]
\centering	
% only \linewidth for double col
\includegraphics[width=\linewidth ]{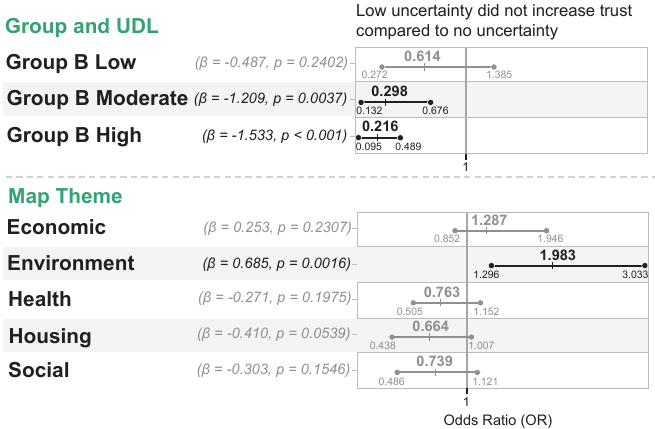}
\caption{CLMM results for H2b: Odds ratios (with 95\% confidence intervals) for the combined Group $\times$ Uncertainty Dominance Level (UDL) predictor and Map Theme, shown relative to the reference levels (Group A and Crime, not shown). The coefficient $\beta$ reflects each predictor’s log-odds effect, and the plotted odds ratios represent exp($\beta$). Statistically significant values at $\alpha$ = 0.05 are shown in \textbf{bold}.}
\label{fig:table_H2b_results}
\Description{This figure presents CLMM results for H2b using odds ratios and 95\% confidence intervals. The top panel shows the combined Group × Uncertainty Dominance Level (UDL) predictor, comparing Group A (no uncertainty) with Group B Low, Moderate, and High. The odds ratios illustrate that low-uncertainty maps do not differ significantly from no-uncertainty maps, while moderate- and high-uncertainty levels are associated with substantially lower odds of assigning higher trust ratings. The bottom panel displays odds ratios for the five non-reference map themes (reference: Crime), showing their individual effects on trust. Statistically significant values are indicated in bold. The visualization conveys how increasing uncertainty dominance reduces trust while low uncertainty remains statistically indistinguishable from no uncertainty.}
\end{figure}

As shown in Fig.~\ref{fig:table_H2b_results}, \textbf{Group B\_Low did not differ significantly from Group A in trust ratings.} The estimated coefficient was –0.487, corresponding to an odds ratio of 0.614 (p = 0.2402). This suggests that the maps dominated by low uncertainty were slightly less likely to receive higher trust ratings than maps without uncertainty, but the difference was not statistically significant. Hence, these findings provide no evidence that low uncertainty visualization increases trust. \textbf{We therefore reject H2b.}

% \begin{figure}[b]
% \centering	
% % only \linewidth for double col
% \includegraphics[width=0.6\linewidth ]{table/interaction_group_theme.pdf}
% \caption{The model includes three sets of predictors: Group (reference = Group A), Map Theme (reference = Crime), and their interaction terms (Group:Theme), which capture how the effect of each theme differs between groups. Estimates indicate the direction and magnitude of effects on the log-odds of assigning higher trust ratings, while Odds Ratios reflect the corresponding multiplicative change in odds. Significant effects (p < .05) are shown in bold.}
% \label{fig:table_interaction_group_theme}
% \Description{The table presents three sets of predictors and corresponding estimates, odds ratios, and p-values. The predictor sets include Group (with Group A as the reference), Map Theme (with the Crime theme as the reference), and their interaction terms.}
% \end{figure}

\subsection{Additional Analysis - MAPTRUST Adjectives}
\label{subsection4.2:additional_analysis}

% \begin{figure}[b]
% \centering	
% % only \linewidth for double col
% \includegraphics[width=0.6\linewidth ]{table/interaction_group_theme.pdf}
% \caption{The model includes three sets of predictors: Group (reference = Group A), Map Theme (reference = Crime), and their interaction terms (Group:Theme), which capture how the effect of each theme differs between groups. Estimates indicate the direction and magnitude of effects on the log-odds of assigning higher trust ratings, while Odds Ratios reflect the corresponding multiplicative change in odds. Significant effects (p < .05) are shown in bold.}
% \label{fig:table_interaction_group_theme}
% \Description{The table presents three sets of predictors and corresponding estimates, odds ratios, and p-values. The predictor sets include Group (with Group A as the reference), Map Theme (with the Crime theme as the reference), and their interaction terms.}
% \end{figure}

\begin{figure}[t]
\centering	
% only \linewidth for double col
\includegraphics[width=0.9\linewidth ]{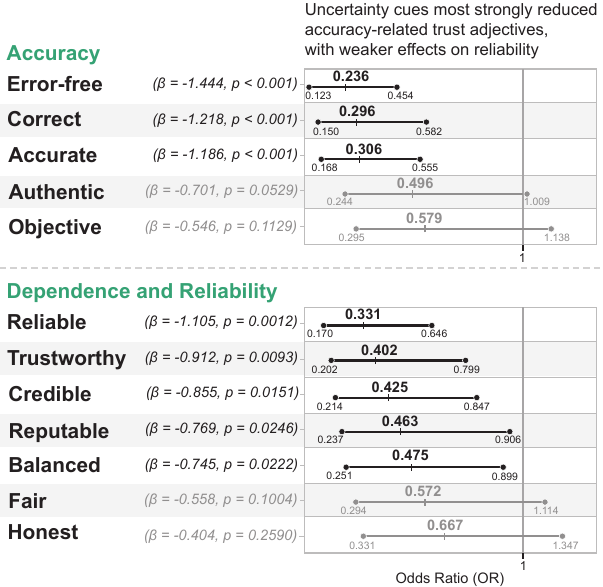} % was matrust_adjective_results.pdf
\caption{Odds ratios (with 95\% confidence intervals) for each MAPTRUST adjective, divided into two dimensions of trust: Accuracy and Dependence/Reliability. The coefficient $\beta$ reflects the estimated log-odds effect for Group B relative to Group A. Values significant at the $\alpha$ = 0.05 level are shown in \textbf{bold}.}
\label{fig:table_maptrust_adjective_results}
\Description{This figure presents a set of horizontal confidence-interval plots showing odds ratios for each MAPTRUST adjective, comparing Group B (uncertainty visualized) to Group A (no uncertainty). The adjectives are organized into two sections that correspond to the MAPTRUST dimensions: Accuracy (Error-free, Correct, Accurate, Authentic, Objective) and Dependence/Reliability (Reliable, Trustworthy, Credible, Reputable, Balanced, Fair, Honest). For each adjective, the plot shows the estimated odds ratio with a 95\% confidence interval, alongside the corresponding coefficient and p-value. Odds ratios below 1 indicate lower odds of assigning higher trust ratings under uncertainty. Statistical significance at alpha = 0.05 is indicated in bold in the accompanying text labels. The visual layout groups adjectives by dimension to highlight contrasts across trust components.}
\end{figure}

In addition to the primary hypothesis tests, we conducted exploratory analyses to examine whether individual trust adjectives from the MAPTRUST scale exhibited distinct patterns in response to uncertainty visualization. Unlike previous models that used overall trust as the dependent variable, each of the twelve MAPTRUST adjectives was modeled separately. For each model, the fixed effects included Group and Map Theme, consistent with earlier analyses. The odds ratios and coefficient estimates reported in Fig.~\ref{fig:table_maptrust_adjective_results} reflect the effect of Group B relative to Group A for each adjective.

Results reveal a clear pattern. Adjectives associated with the \textbf{\inputvariable{Accuracy}} dimension of trust—such as Error-free, Correct, and Accurate—exhibited the strongest and most statistically significant negative effects for Group B. In contrast, adjectives tied to the \textbf{\inputvariable{Dependability and Reliability}} dimension (see adjective classification in Section~\ref{subsection2.1:trust_in_maps}; also see Appendix~\ref{appendixA2:maptrust_adjectives}) showed relatively weaker differences between groups. Most notably, Fair and Honest showed no statistically significant differences at all.

These results suggest that\textbf{ uncertainty visualization may selectively diminish perceptions of accuracy, while having a lesser impact on judgments related to reliability.} This does not imply that perceptions of dependability were unaffected—only that the magnitude of group difference was smaller than for accuracy-related evaluations.

\section{Discussion}
\label{section5:discussion}

\subsection{Effect of Uncertainty Visualization on Trust}
\label{subsection5.1:uncertainty_visualization_and_trust}

As per the results on H1 and H2a, we see that \textbf{uncertainty visualization has a measurable impact on trust in thematic maps, and that the negative effect of uncertainty grows with its magnitude.} Maps dominated by moderate or high uncertainty showed progressively larger declines in trust, indicating that participants were sensitive to the degree of visual uncertainty presented. This pattern may stem from how increasing uncertainty visually disrupts the clarity or perceived reliability of the underlying data, leading users to question the map’s validity or usefulness~\cite{hullman2019authors}.

In addition, the findings for H2b showed that maps with low UDL did not significantly differ from maps without uncertainty with respect to trust ratings. In other words, \textbf{low uncertainty visualization appears trust-neutral rather than trust-boosting or trust-reducing}. This suggests that while small amounts of uncertainty do not enhance trust, they also do not significantly undermine it, unlike moderate or high levels where trust is clearly reduced.

These results emphasize that the impact of uncertainty visualization on trust is not binary. Instead, it appears to operate along a gradient, where \textbf{moderate and high levels of uncertainty dominance produce measurable decreases in trust}, but low levels do not significantly shift trust perceptions compared to baseline (no uncertainty). This may be because low levels of uncertainty are either too subtle to register as alarming or are cognitively discounted by users as negligible noise, thereby failing to trigger a strong response in either direction. This gradient pattern complements Padilla et al.’s~\cite{padilla2022multiple} finding that greater perceptual load in uncertainty displays can suppress trust. However, in our case, the mechanism relates not to visual complexity but to the stronger visual salience of uncertainty.

Taken together, these findings suggest that uncertainty visualization does not uniformly erode trust; rather, its impact depends on how visibly uncertainty is conveyed. Low-magnitude uncertainty cues appeared to register but did not meaningfully shift trust, indicating that readers may treat small uncertainty as tolerable noise rather than as a warning signal. Instead of implying that uncertainty should be minimized or hidden, this reinforces that the salience and framing of uncertainty matter—readers react more strongly when uncertainty appears visually dominant or severe.

As reviewed in Section~\ref{subsection2.3:geospatial_uncertainty}, prior work shows that uncertainty encodings can reshape how readers interpret and extract takeaways from maps~\cite{correll2018value, ndlovu2023taken}. Our findings extend this perspective by demonstrating that such encodings also modulate readers’ trust in the map, not only their interpretation of its content.

Our results are based on a deliberately extreme experimental condition in which high-magnitude uncertainty meant that most values on the map were depicted as highly uncertain. Such situations are unlikely to arise in practice, where a cartographer might instead choose not to map highly uncertain data at all. Therefore, these findings motivate follow-up research testing whether similar trust effects emerge when uncertainty is embedded in more realistic settings where uncertain values are sparse, localized, or accompanied by contextual explanation.

\subsection{Adjective-Level Patterns in Trust Judgments}
\label{subsection5.2:adjective_level_discussions}

Section~\ref{subsection4.2:additional_analysis} showed that the two dimensions of trust were not equally affected by uncertainty visualization. Adjectives tied to the accuracy dimension showed the strongest and most statistically significant negative effects when uncertainty was visualized. \textbf{This suggests that seeing uncertainty cues led participants to question the data’s precision and correctness}, likely due to the visual emphasis on variability, missing information, or imprecision.

In contrast, adjectives related to dependability and reliability showed weaker effects. Notably, terms like Fair and Honest did not differ significantly between conditions. This indicates that \textbf{readers may not interpret visualized uncertainty as a reflection of the mapmaker’s intent or ethics.}

These findings highlight the multidimensional nature of trust. While uncertainty visualization erodes certain dimensions more strongly (particularly accuracy-related judgments), it does not lead to uniform distrust. Participants distinguish between uncertainty in data quality and assumptions about the mapmaker's intent or ethical integrity. This distinction also aligns with the theoretical separation between cognitive and affective trust discussed in Section~\ref{subsection2.1:trust_in_maps}, particularly as articulated in the Vistrust framework~\cite{elhamdadi2023vistrust}. Cognitive components of trust (e.g., perceived accuracy) tend to be more sensitive to uncertainty cues than affective components (e.g., perceived benevolence), a pattern that mirrors the adjective-level effects observed in our study.

\section{Limitations and Future Work}
\label{section6:limitations_future_work}

While our study offers new insights into how uncertainty visualization affects trust in thematic maps, several limitations merit discussion and offer promising directions for future research. First, the study focused on a specific set of thematic topics—social, environment, housing, crime, health, and economic—at the U.S. state level. Although these themes were chosen for their prevalence in public-facing maps, they may not generalize to maps at finer geographic scales (e.g., counties or neighborhoods) or to more controversial or technical topics such as migration, political polling, or biodiversity. Future studies should explore whether the effects of uncertainty visualization differ across geographic granularity and thematic domains, particularly in contexts that evoke strong prior beliefs or domain-specific expertise, and/or in domains where intent and ethics are more likely to be questioned.

Related to this point, we observed an exploratory pattern suggesting that uncertainty visualization may reduce differences in trust across map themes: when uncertainty was absent, trust ratings varied more strongly by topic, whereas these differences appeared to be dampened once uncertainty was visualized. A small exception was the Environment theme, which showed slightly higher trust under uncertainty; however, because the study was not designed to test explanations for theme-specific differences, we refrain from inferring mechanisms. Full model details and results are provided in our supplementary exploratory analysis (Appendix~\ref{appendixA3:map_theme_exploratory}), and we report these patterns as descriptive observations that may motivate future research on how thematic context interacts with uncertainty in map-based communication.

Second, we used a single uncertainty visualization technique based on established visual encodings of attribute fuzziness. This choice allowed us to maintain experimental control and minimize confounds, but it also means our results may not generalize to other types of uncertainty representations, such as confidence intervals, animated variability, hatched overlays, or multiple model forecasts. Future work should systematically compare these design alternatives to assess whether particular visual strategies are more effective in maintaining trust without overwhelming users.

Third, although our maps used real attribute data, the uncertainty values were synthetic and assigned in controlled proportions to create clearly distinguishable low, moderate, and high uncertainty conditions based on our experimental needs. This approach was necessary to avoid confounds from naturally occurring spatial clustering in real-world uncertainty patterns. However, this design choice meant that participants relied on qualitative descriptors (e.g., “smaller or more variable samples”) rather than precise numeric values, which may have left room for varied interpretations among non-expert readers. Prior work indicates that numeric expressions of uncertainty (e.g., margins of error) can anchor interpretation more effectively than ordinal labels~\cite{van2020effects}. Future studies should therefore test how trust is influenced when uncertainty is: (a) real rather than synthetic, (b) numeric rather than ordinal, or (c) accompanied by provenance indicators explaining how uncertainty was computed.

Finally, our trust measurements relied on the MAPTRUST adjective scale—a validated and multidimensional instrument—but they were still self-reported. This raises the possibility that participants’ trust ratings reflect momentary impressions rather than deeper trust that would influence real-world action. More broadly, trust is a latent psychological construct that can only be inferred indirectly, meaning that any single measurement approach—whether self-report or behavioral—may capture it only partially, especially when interpreting responses to uncertainty. Future work should therefore combine self-report, behavioral, and longitudinal methods to better understand how trust in uncertain maps forms, shifts, and stabilizes over time.

Together, these limitations highlight the need for broader evaluations across themes, scales, and uncertainty designs. As the use of thematic maps continues to grow in public communication, understanding how to design for trust—without oversimplifying or obscuring uncertainty—remains a critical research challenge.

\section{Conclusion}

In an information landscape increasingly shaped by data visualizations, establishing how design choices affect public trust is essential—particularly for widely circulated tools like thematic maps. This study advances the empirical understanding of how visualizing uncertainty influences trust in thematic maps among non-expert audiences. Unlike prior work that either focused on expert users or treated trust as a secondary outcome, our study explicitly measured trust using a validated multidimensional scale and systematically manipulated both the presence and level of uncertainty visualization. The results reveal that while low uncertainty may not erode trust, higher levels significantly reduce perceived trustworthiness—at least when uncertainty is depicted using intrinsic, coincident, static representations such as symbol fuzziness—especially in relation to a map’s perceived accuracy. Together, these findings highlight the importance of designing uncertainty representations that are transparent yet easy to interpret. As thematic maps continue to play a critical role in public decision-making and discourse, our work offers timely, evidence-based guidance for fostering both honesty and credibility in geovisual storytelling.

% \begin{acks}
\begin{acks}
This work is supported in part by the Natural Sciences and Engineering Research Council of Canada (NSERC) Discovery Grant \#RGPIN-2020-03966.
\end{acks}
% \end{acks}

\bibliographystyle{ACM-Reference-Format}
\bibliography{sample-base}

@String{Computing = "Computing" }

@String{Computer = "{IEEE} Computer" }

@String{Springer = "Springer-Verlag" }

@article{kinkeldey2017evaluating,
  title={Evaluating the effect of visually represented geodata uncertainty on decision-making: systematic review, lessons learned, and recommendations},
  author={Kinkeldey, Christoph and MacEachren, Alan M and Riveiro, Maria and Schiewe, Jochen},
  journal={Cartography and Geographic Information Science},
  volume={44},
  number={1},
  pages={1--21},
  year={2017},
  publisher={Taylor \& Francis}
}

@article{hullman2019authors,
  title={Why authors don't visualize uncertainty},
  author={Hullman, Jessica},
  journal={IEEE Transactions on Visualization and Computer Graphics},
  volume={26},
  number={1},
  pages={130--139},
  year={2019},
  publisher={IEEE}
}

@article{prestby2023trust,
  title={Trust in maps: what we know and what we need to know},
  author={Prestby, Timothy J},
  journal={Cartography and Geographic Information Science},
  pages={1--18},
  year={2023},
  publisher={Taylor \& Francis}
}

@inproceedings{maceachren2015visual,
  title={Visual analytics and uncertainty: Its not about the data},
  author={MacEachren, Alan M},
  year={2015},
  booktitle={Workshop on Visual Analytics of the Eurographics/IEEE VGTC Conference on Visualization}
}

@article{maceachren2005visualizing,
  title={Visualizing geospatial information uncertainty: What we know and what we need to know},
  author={MacEachren, Alan M and Robinson, Anthony and Hopper, Susan and Gardner, Steven and Murray, Robert and Gahegan, Mark and Hetzler, Elisabeth},
  journal={Cartography and Geographic Information Science},
  volume={32},
  number={3},
  pages={139--160},
  year={2005},
  publisher={Taylor \& Francis}
}

@article{kinkeldey2014assess,
  title={How to assess visual communication of uncertainty? {A} systematic review of geospatial uncertainty visualisation user studies},
  author={Kinkeldey, Christoph and MacEachren, Alan M and Schiewe, Jochen},
  journal={The Cartographic Journal},
  volume={51},
  number={4},
  pages={372--386},
  year={2014},
  publisher={Taylor \& Francis}
}

@article{maceachren2012visual,
  title={Visual semiotics \& uncertainty visualization: An empirical study},
  author={MacEachren, Alan M and Roth, Robert E and O'Brien, James and Li, Bonan and Swingley, Derek and Gahegan, Mark},
  journal={IEEE Transactions on Visualization and Computer Graphics},
  volume={18},
  number={12},
  pages={2496--2505},
  year={2012},
  publisher={IEEE}
}

@article{schofield2006survey,
  title={Survey sampling},
  author={Schofield, William},
  journal={Data Collection and Analysis},
  volume={2},
  pages={332},
  year={2006},
  publisher={Sage London, UK}
}

@inproceedings{fan2024understanding,
  title={Understanding reader takeaways in thematic maps under varying text, detail, and spatial autocorrelation},
  author={Fan, Arlen and Lei, Fan and Mancenido, Michelle and Maceachren, Alan M and Maciejewski, Ross},
  booktitle={Proceedings of the ACM Conference on Human Factors in Computing Systems},
  pages={1--17},
  year={2024}
}

@article{jenks1971error,
  title={Error on choroplethic maps: definition, measurement, reduction},
  author={Jenks, George F and Caspall, Fred C},
  journal={Annals of the Association of American Geographers},
  volume={61},
  number={2},
  pages={217--244},
  year={1971},
  publisher={Taylor \& Francis}
}

@book{dent1999cartography,
  title={Cartography: Thematic Map Design},
  author={Dent, Borden D},
  year={1999},
  publisher={WCB/McGraw-Hill}
}

@techreport{snyder1982map,
  title={Map projections used by the {U.S.} {G}eological {S}urvey},
  author={Snyder, John Parr},
  year={1982},
  institution={US Government Printing Office}
}

@article{ruginski2016non,
  title={Non-expert interpretations of hurricane forecast uncertainty visualizations},
  author={Ruginski, Ian T and Boone, Alexander P and Padilla, Lace M and Liu, Le and Heydari, Nahal and Kramer, Heidi S and Hegarty, Mary and Thompson, William B and House, Donald H and Creem-Regehr, Sarah H},
  journal={Spatial Cognition \& Computation},
  volume={16},
  number={2},
  pages={154--172},
  year={2016},
  publisher={Taylor \& Francis}
}

@article{seipel2017color,
  title={Color map design for visualization in flood risk assessment},
  author={Seipel, Stefan and Lim, Nancy Joy},
  journal={International Journal of Geographical Information Science},
  volume={31},
  number={11},
  pages={2286--2309},
  year={2017},
  publisher={Taylor \& Francis}
}

@inproceedings{thomson2005typology,
  title={A typology for visualizing uncertainty},
  author={Thomson, Judi and Hetzler, Elizabeth and MacEachren, Alan and Gahegan, Mark and Pavel, Misha},
  booktitle={Visualization and Data Analysis},
  volume={5669},
  pages={146--157},
  year={2005}
}

@book{slocum2022thematic,
  title={Thematic Cartography and Geovisualization},
  author={Slocum, Terry A and McMaster, Robert B and Kessler, Fritz C and Howard, Hugh H},
  year={2022},
  publisher={CRC Press}
}

@book{brewer2016designing,
  title={Designing Better Maps: A Guide for GIS Users, 2nd Edition},
  author={Brewer, Cynthia},
  year={2016},
  publisher={ESRI press}
}

@book{montgomery2017design,
  title={Design and Analysis of Experiments},
  author={Montgomery, Douglas C},
  year={2017},
  publisher={John Wiley \& Sons}
}

@ARTICLE{10273434,
  author={Lei, Fan and Fan, Arlen and MacEachren, Alan M. and Maciejewski, Ross},
  journal={IEEE Transactions on Visualization and Computer Graphics}, 
  title={GeoLinter: A Linting Framework for Choropleth Maps}, 
  year={2023},
  volume={},
  number={},
  pages={1-16}}

@inproceedings{he2011visualize,
  title={To visualize spatial data using thematic maps combined with infographics},
  author={He, Manli and Tang, Xi and Huang, Yuming},
  booktitle={International Conference on Geoinformatics},
  pages={1--5},
  year={2011}
}

@article{roth2021cartographic,
  title={Cartographic design as visual storytelling: Synthesis and review of map-based narratives, genres, and tropes},
  author={Roth, Robert E},
  journal={The Cartographic Journal},
  volume={58},
  number={1},
  pages={83--114},
  year={2021},
  publisher={Taylor \& Francis}
}

@article{song2022visual,
  title={Visual storytelling with maps: An empirical study on story map themes and narrative elements, visual storytelling genres and tropes, and individual audience differences},
  author={Song, Zihan and Roth, Robert E and Houtman, Lily and Prestby, Timothy and Iverson, Alicia and Gao, Song},
  journal={Cartographic Perspectives},
  number={100},
  pages={10--44},
  year={2022}
}

@article{raposo2020change,
  title={A change of theme: The role of generalization in thematic mapping},
  author={Raposo, Paulo and Touya, Guillaume and Bereuter, Pia},
  journal={ISPRS International Journal of Geo-Information},
  volume={9},
  number={6},
  pages={371},
  year={2020},
  publisher={MDPI}
}

@article{mocnik2020epidemics,
  title={Epidemics and pandemics in maps--{T}he case of {COVID}-19},
  author={Mocnik, Franz-Benjamin and Raposo, Paulo and Feringa, Wim and Kraak, Menno-Jan and K{\"o}bben, Barend},
  journal={Journal of Maps},
  volume={16},
  number={1},
  pages={144--152},
  year={2020},
  publisher={Taylor \& Francis}
}

@article{juergens2020trustworthy,
  title={Trustworthy {COVID}-19 mapping: Geo-spatial data literacy aspects of choropleth maps},
  author={Juergens, Carsten},
  journal={KN-Journal of Cartography and Geographic Information},
  volume={70},
  number={4},
  pages={155--161},
  year={2020},
  publisher={Springer}
}

@article{joslyn2012uncertainty,
  title={Uncertainty forecasts improve weather-related decisions and attenuate the effects of forecast error.},
  author={Joslyn, Susan L and LeClerc, Jared E},
  journal={Journal of Experimental Psychology: Applied},
  volume={18},
  number={1},
  pages={126},
  year={2012},
  publisher={American Psychological Association}
}

@article{savelli2013advantages,
  title={The advantages of predictive interval forecasts for non-expert users and the impact of visualizations},
  author={Savelli, Sonia and Joslyn, Susan},
  journal={Applied Cognitive Psychology},
  volume={27},
  number={4},
  pages={527--541},
  year={2013},
  publisher={Wiley Online Library}
}

@inproceedings{maceachren1995mapping,
  title={Mapping health statistics: Representing data reliability},
  author={MacEachren, Alan M and Brewer, Cynthia A and Pickle, Linda W},
  booktitle={Proceedings of the International Cartographic Conference},
  pages={311--319},
  year={1995}
}

@article{huang2019exploring,
  title={Exploring the sensitivity of choropleths under attribute uncertainty},
  author={Huang, Zhaosong and Lu, Yafeng and Mack, Elizabeth A and Chen, Wei and Maciejewski, Ross},
  journal={IEEE Transactions on Visualization and Computer Graphics},
  volume={26},
  number={8},
  pages={2576--2590},
  year={2019},
  publisher={IEEE}
}

@article{karduni2020bayesian,
  title={A bayesian cognition approach for belief updating of correlation judgement through uncertainty visualizations},
  author={Karduni, Alireza and Markant, Douglas and Wesslen, Ryan and Dou, Wenwen},
  journal={IEEE Transactions on Visualization and Computer Graphics},
  volume={27},
  number={2},
  pages={978--988},
  year={2020},
  publisher={IEEE}
}

@article{propen2007visual,
  title={Visual communication and the map: How maps as visual objects convey meaning in specific contexts},
  author={Propen, Amy},
  journal={Technical Communication Quarterly},
  volume={16},
  number={2},
  pages={233--254},
  year={2007},
  publisher={Taylor \& Francis}
}

@article{spielman2014patterns,
  title={Patterns and causes of uncertainty in the American Community Survey},
  author={Spielman, Seth E and Folch, David and Nagle, Nicholas},
  journal={Applied Geography},
  volume={46},
  pages={147--157},
  year={2014},
  publisher={Elsevier}
}

@article{calfano2022bad,
  title={Bad impressions: How journalists as “storytellers” diminish public confidence in media},
  author={Calfano, Brian and Blevins, Jeffrey Layne and Straka, Alexis},
  journal={Journal of Broadcasting \& Electronic Media},
  volume={66},
  number={1},
  pages={176--199},
  year={2022},
  publisher={Taylor \& Francis}
}

@article{prestby2024measuring,
  title={Measuring trust in maps: development and evaluation of the MAPTRUST scale},
  author={Prestby, Timothy J},
  journal={International Journal of Geographical Information Science},
  volume={38},
  number={10},
  pages={2083--2107},
  year={2024},
  publisher={Taylor \& Francis}
}

@phdthesis{prestby2025evaluating,
  title={Evaluating Trust in Cartographic Representations: Theory, Methods, and Design},
  author={Prestby, Tim J},
  year={2025},
  school={The Pennsylvania State University}
}

@misc{us_census_acs,
  author       = {{U.S. Census Bureau}},
  title        = {American Community Survey (ACS)},
  year         = {2024},
  url          = {https://www.census.gov/programs-surveys/acs}
}

@misc{fbi_crime_data_explorer,
  author       = {{Federal Bureau of Investigation}},
  title        = {Crime Data Explorer},
  year         = {2024},
  url          = {https://cde.ucr.cjis.gov/}
}

@misc{cdc_data,
  author       = {{Centers for Disease Control and Prevention}},
  title        = {CDC Public Health Data},
  year         = {2024},
  url          = {https://www.cdc.gov/places/index.html}
}

@misc{epa_data,
  author       = {{U.S. Environmental Protection Agency}},
  title        = {EPA Environmental Data},
  year         = {2024},
  url          = {https://www.epa.gov/aqs}
}

@misc{uscb_econ_geo_levels,
  author       = {{U.S. Census Bureau}},
  title        = {Geographic Levels},
  year         = {2022},
  url          = {https://www.census.gov/programs-surveys/economic-census/guidance-geographies/levels.html},
  note         = {Accessed: 2025-03-22}
}

@misc{prolific2025,
  author       = {{Prolific}},
  title        = {{Prolific Participant Recruitment Platform}},
  year         = {2025},
  howpublished = {\url{https://www.prolific.com}},
  note         = {Version used: September 2025. Company based in London, UK. First released in 2014}
}

@article{krosnick1987evaluation,
  title={An evaluation of a cognitive theory of response-order effects in survey measurement},
  author={Krosnick, Jon A and Alwin, Duane F},
  journal={Public Opinion Quarterly},
  volume={51},
  number={2},
  pages={201--219},
  year={1987},
  publisher={Oxford University Press}
}

@article{mccullagh1980regression,
  title={Regression models for ordinal data},
  author={McCullagh, Peter},
  journal={Journal of the Royal Statistical Society: Series B (Methodological)},
  volume={42},
  number={2},
  pages={109--127},
  year={1980},
  publisher={Wiley Online Library}
}

@book{agresti2010analysis,
  title={Analysis of Ordinal Categorical Data},
  author={Agresti, Alan},
  year={2010},
  publisher={John Wiley \& Sons}
}

@book{greene2010modeling,
  title={Modeling Ordered Choices: A Primer},
  author={Greene, William H and Hensher, David A},
  year={2010},
  publisher={Cambridge University Press}
}

@article{christensen2018cumulative,
  title={Cumulative link models for ordinal regression with the R package ordinal},
  author={Christensen, Rune Haubo B},
  journal={Journal of Statistical Software},
  volume={35},
  pages={1--46},
  year={2018}
}

@article{griffin2020trustworthy,
  title={Trustworthy maps},
  author={Griffin, Amy L},
  journal={Journal of Spatial Information Science},
  volume={2020},
  number={20},
  pages={5--19},
  year={2020}
}

@article{padilla2022multiple,
  title={Multiple forecast visualizations (mfvs): Trade-offs in trust and performance in multiple {COVID}-19 forecast visualizations},
  author={Padilla, Lace and Fygenson, Racquel and Castro, Spencer C and Bertini, Enrico},
  journal={IEEE Transactions on Visualization and Computer Graphics},
  volume={29},
  number={1},
  pages={12--22},
  year={2022},
  publisher={IEEE}
}

@article{schiewe2013vertrauen,
  title={Vertrauen im rahmen der nutzung von karten},
  author={Schiewe, Jochen and Schweer, Martin KW},
  journal={KN-Journal of Cartography and Geographic Information},
  volume={63},
  number={2},
  pages={59--66},
  year={2013},
  publisher={Springer}
}

@article{mcgranaghan1999web,
  title={The Web, cartography and trust},
  author={McGranaghan, Matthew},
  journal={Cartographic Perspectives},
  number={32},
  pages={3--5},
  year={1999}
}

@incollection{skarlatidou2011understanding,
  title={Understanding the influence of specific Web GIS attributes in the formation of non-experts’ trust perceptions},
  author={Skarlatidou, Artemis and Wardlaw, Jessica and Haklay, Muki and Cheng, Tao},
  booktitle={Advances in Cartography and GIScience. Volume 1: Selection from ICC 2011, Paris},
  pages={219--238},
  year={2011},
  publisher={Springer}
}

@article{skarlatidou2013guidelines,
  title={Guidelines for trust interface design for public engagement Web GIS},
  author={Skarlatidou, Artemis and Cheng, Tao and Haklay, Muki},
  journal={International Journal of Geographical Information Science},
  volume={27},
  number={8},
  pages={1668--1687},
  year={2013},
  publisher={Taylor \& Francis}
}

@inproceedings{xiong2019examining,
  title={Examining the components of trust in map-based visualizations},
  author={Xiong, Cindy and Padilla, Lace and Grayson, Kent and Franconeri, Steven},
  booktitle={Workshop on Trustworthy Visualization of the Eurographics/IEEE VGTC Conference on Visualization},
  pages={19--23},
  year={2019},
  organization={The Eurographics Association}
}

@article{kubler2020against,
  title={Against all odds: Multicriteria decision making with hazard prediction maps depicting uncertainty},
  author={K{\"u}bler, Isabella and Richter, Kai-Florian and Fabrikant, Sara Irina},
  journal={Annals of the American Association of Geographers},
  volume={110},
  number={3},
  pages={661--683},
  year={2020},
  publisher={Taylor \& Francis}
}

@misc{wiki_midatlantic,
  title        = {Mid‑Atlantic (United States)},
  howpublished = {\url{https://en.wikipedia.org/wiki/Mid-Atlantic_(United_States)}},
  note         = {Accessed: 2025‑09‑11},
  year         = {2025}
}

@article{li2018communicating,
  title={Communicating data: interactive infographics, scientific data and credibility},
  author={Li, Nan TTU and Brossard, Dominique and Scheufele, Dietram A and Wilson, Paul H and Rose, Kathleen M},
  journal={Journal of Science and Communication},
  year={2018}
}

@article{link2021credibility,
  title={Credibility and enjoyment through data? Effects of statistical information and data visualizations on message credibility and reading experience},
  author={Link, Elena and Henke, Jakob and M{\"o}hring, Wiebke},
  journal={Journalism Studies},
  volume={22},
  number={5},
  pages={575--594},
  year={2021},
  publisher={Taylor \& Francis}
}

@inproceedings{kong2019trust,
  title={Trust and recall of information across varying degrees of title-visualization misalignment},
  author={Kong, Ha-Kyung and Liu, Zhicheng and Karahalios, Karrie},
  booktitle={Proceedings of the ACM Conference on Human Factors in Computing Systems},
  pages={1--13},
  year={2019}
}

@article{searle1980population,
  title={Population marginal means in the linear model: an alternative to least squares means},
  author={Searle, Shayle R and Speed, F Michael and Milliken, George A},
  journal={The American Statistician},
  volume={34},
  number={4},
  pages={216--221},
  year={1980},
  publisher={Taylor \& Francis}
}

@article{elhamdadi2023vistrust,
  title={Vistrust: a multidimensional framework and empirical study of trust in data visualizations},
  author={Elhamdadi, Hamza and Stefkovics, Adam and Beyer, Johanna and Moerth, Eric and Pfister, Hanspeter and Bearfield, Cindy Xiong and Nobre, Carolina},
  journal={IEEE Transactions on Visualization and Computer Graphics},
  volume={30},
  number={1},
  pages={348--358},
  year={2023},
  publisher={IEEE}
}

@inproceedings{peck2019data,
  title={Data is personal: Attitudes and perceptions of data visualization in rural pennsylvania},
  author={Peck, Evan M and Ayuso, Sofia E and El-Etr, Omar},
  booktitle={Proceedings of the ACM Conference on Human Factors in Computing Systems},
  pages={1--12},
  year={2019}
}

@inproceedings{correll2018value,
  title={Value-suppressing uncertainty palettes},
  author={Correll, Michael and Moritz, Dominik and Heer, Jeffrey},
  booktitle={Proceedings of the ACM Conference on Human Factors in Computing Systems},
  pages={1--11},
  year={2018}
}

@inproceedings{ndlovu2023taken,
  title={Taken By Surprise? Evaluating how Bayesian Surprise \& Suppression Influences Peoples’ Takeaways in Map Visualizations},
  author={Ndlovu, Akim and Shrestha, Hilson and Harrison, Lane T},
  booktitle={IEEE Visualization and Visual Analytics},
  pages={136--140},
  year={2023}
}

@article{van2020effects,
  title={The effects of communicating uncertainty on public trust in facts and numbers},
  author={Van Der Bles, Anne Marthe and van der Linden, Sander and Freeman, Alexandra LJ and Spiegelhalter, David J},
  journal={Proceedings of the National Academy of Sciences},
  volume={117},
  number={14},
  pages={7672--7683},
  year={2020},
  publisher={National Academy of Sciences}
}

@article{fox2025quantifying,
  title={Quantifying Visualization Vibes: Measuring Socio-Indexicality at Scale},
  author={Fox, Amy Rae and Morgenstern, Michelle and Jones, Graham M and Satyanarayan, Arvind},
  journal={IEEE Transactions on Visualization and Computer Graphics},
  year={2025}
}

@article{morgenstern2025visualization,
  title={Visualization Vibes: The Socio-Indexical Function of Visualization Design},
  author={Morgenstern, Michelle and Fox, Amy Rae and Jones, Graham M and Satyanarayan, Arvind},
  journal={IEEE Transactions on Visualization and Computer Graphics},
  year={2025}
}

\appendix

\section{Appendix}

\subsection{U.S. Region Definitions and Groupings}
\label{appendixA1:region_groupings}

This study followed the U.S. Census Bureau’s four-region schema~\cite{uscb_econ_geo_levels} to ensure balanced geographic representation when assigning uncertainty levels to states. The four regions used in our experiment are:

\begin{itemize}
    \item \textbf{Northeast: }Includes states in the northeastern part of the country, often associated with early industrialization and high urban density.
    \item \textbf{Midwest:} Comprises states in the north-central U.S., characterized by agricultural and manufacturing economies.
    \item \textbf{South:} Encompasses southeastern and south-central states, often grouped together due to shared cultural and economic characteristics.
    \item \textbf{West:} Covers states west of the Rocky Mountains, generally known for expansive geography and diverse environmental conditions.
\end{itemize}

The classification of each of the 48 contiguous U.S. states into one of the four regions is provided in Table~\ref{tab:us-region-groupings}. While these groupings largely follow the official U.S. Census Bureau classification, we made a deliberate adjustment: Maryland and Delaware, which are officially categorized under the South, were reassigned to the Northeast for this study. This modification was necessary to address an initial imbalance—under the original schema, the Northeast contained only 9 states, fewer than the other regions. The adjustment helped improve regional parity during uncertainty level assignment. Additionally, this change is supported by common public perception, as Maryland and Delaware are often referred to as Mid-Atlantic states—a subregion typically associated with the Northeastern U.S. in cultural and geographic contexts~\cite{wiki_midatlantic}.

 \begin{table}[]
\centering
\begin{tabular}{|l|l|l|}
\hline
\textbf{State Name} & \textbf{Abbreviation} & \textbf{Region}    \\ \hline
Alabama             & AL           & South     \\ \hline
Arkansas            & AR           & South     \\ \hline
Arizona             & AZ           & West      \\ \hline
California          & CA           & West      \\ \hline
Colorado            & CO           & West      \\ \hline
Connecticut         & CT           & Northeast \\ \hline
Delaware            & DE           & Northeast \\ \hline
Florida             & FL           & South     \\ \hline
Georgia             & GA           & South     \\ \hline
Iowa                & IA           & Midwest   \\ \hline
Idaho               & ID           & West      \\ \hline
Illinois            & IL           & Midwest   \\ \hline
Indiana             & IN           & Midwest   \\ \hline
Kansas              & KS           & Midwest   \\ \hline
Kentucky            & KY           & South     \\ \hline
Louisiana           & LA           & South     \\ \hline
Massachusetts       & MA           & Northeast \\ \hline
Maryland            & MD           & Northeast \\ \hline
Maine               & ME           & Northeast \\ \hline
Michigan            & MI           & Midwest   \\ \hline
Minnesota           & MN           & Midwest   \\ \hline
Missouri            & MO           & Midwest   \\ \hline
Mississippi         & MS           & South     \\ \hline
Montana             & MT           & West      \\ \hline
North Carolina      & NC           & South     \\ \hline
North Dakota        & ND           & Midwest   \\ \hline
Nebraska            & NE           & Midwest   \\ \hline
New Hampshire       & NH           & Northeast \\ \hline
New Jersey          & NJ           & Northeast \\ \hline
New Mexico          & NM           & West      \\ \hline
Nevada              & NV           & West      \\ \hline
New York            & NY           & Northeast \\ \hline
Ohio                & OH           & Midwest   \\ \hline
Oklahoma            & OK           & South     \\ \hline
Oregon              & OR           & West      \\ \hline
Pennsylvania        & PA           & Northeast \\ \hline
Rhode Island        & RI           & Northeast \\ \hline
South Carolina      & SC           & South     \\ \hline
South Dakota        & SD           & Midwest   \\ \hline
Tennessee           & TN           & South     \\ \hline
Texas               & TX           & South     \\ \hline
Utah                & UT           & West      \\ \hline
Virginia            & VA           & South     \\ \hline
Vermont             & VT           & Northeast \\ \hline
Washington          & WA           & West      \\ \hline
Wisconsin           & WI           & Midwest   \\ \hline
West Virginia       & WV           & South     \\ \hline
Wyoming             & WY           & West      \\ \hline
\end{tabular}
\caption{U.S. Region Groupings}
\label{tab:us-region-groupings}
\Description{Table lists the classification of all 48 contiguous U.S. states into four broad geographic regions used in the experiment: West (11 states), Midwest (12), South (14), and Northeast (11). This regional grouping was based largely on the U.S. Census Bureau’s definition but includes a minor adjustment: Maryland and Delaware, originally categorized in the South, were reassigned to the Northeast to improve regional balance during uncertainty assignment. This reclassification is supported by cultural and geographic norms, as these two states are frequently identified as part of the Mid-Atlantic and Northeastern U.S. This balanced grouping ensured that each region had roughly equal representation in the randomized uncertainty dominance assignments applied across map stimuli.}
\end{table}

\subsection{MAPTRUST Adjectives: Dimensions and Definitions}
\label{appendixA2:maptrust_adjectives}

The MAPTRUST framework defines trust in maps as “\textit{the willingness to depend on a map based on the expectation that it is accurate and that acting on its visualized information will not be detrimental.}”~\cite{prestby2024measuring} This definition highlights two key dimensions of trust: the perceived accuracy of the map’s content and the perceived integrity of its creator.

The twelve adjectives used in the MAPTRUST scale were designed to align with one of two overarching dimensions:

\begin{itemize}
    \item \textbf{Accuracy}, which reflects perceptions of the credibility and correctness of the data itself.
    \item \textbf{Dependability and Reliability}, which relate to perceptions of the mapmaker’s intentions, transparency, and the trustworthiness of the source and the map maker.
\end{itemize}

\begin{figure}[h]
\centering	 
% 0.9 for double col
\includegraphics[width=\linewidth]{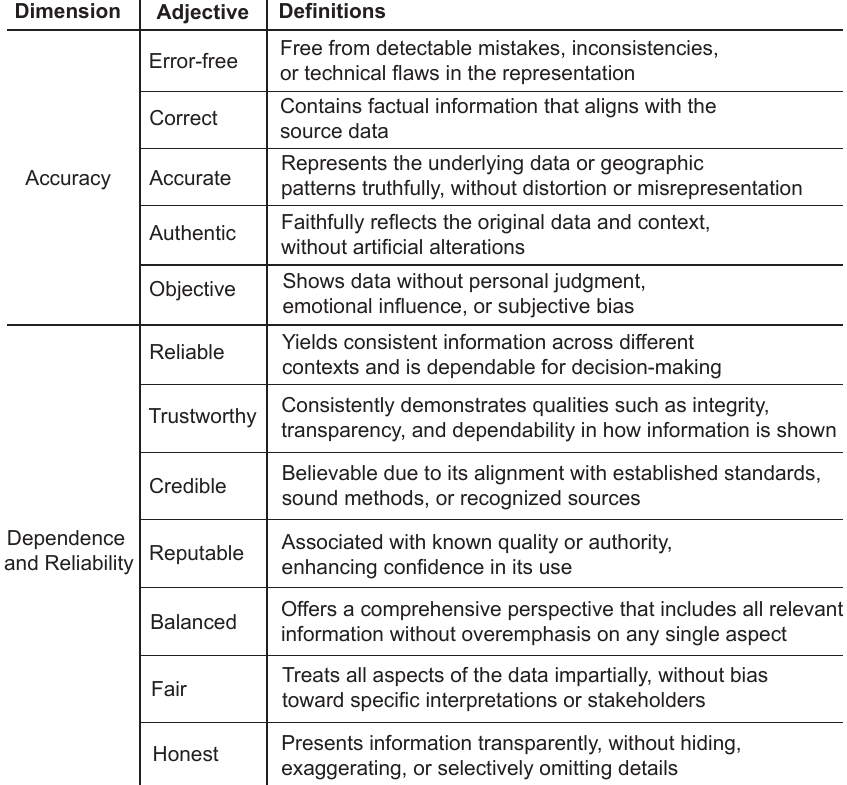}
\caption{Classification of MAPTRUST adjectives into Accuracy and Dependence/Reliability dimensions, along with their definitions.}
\label{fig:table_maptrust_adjective_definitions}
\Description{This figure presents the 12 adjectives used to evaluate trust in thematic maps, as defined in the MAPTRUST framework. Each adjective was associated with one of two overarching dimensions of trust: Accuracy, or Dependence and Reliability. The Accuracy dimension includes adjectives related to the credibility and correctness of the data itself (Error-free, Correct, Accurate, Authentic, and Objective), while the Dependence and Reliability dimension includes those related to the intentions or integrity of the mapmaker and source (Reliable, Trustworthy, Credible, Reputable, Balanced, Fair, and Honest). The definitions were shown to participants on-demand during the experiment and were crafted based on prior literature and validated through expert review. The goal of this classification was to help distinguish which specific aspects of trust were being impacted in participant ratings. This figure supports the Appendix section discussing the operationalization of trust and the alignment of the MAPTRUST adjectives with these two conceptual dimensions.}
\end{figure}

Each adjective was shown to participants during the rating task, with definitions available on demand. These definitions were informed by prior literature across visualization, journalism, and communication~\cite{li2018communicating, link2021credibility, kong2019trust, prestby2024measuring}, and were refined through expert review. Fig.~\ref{fig:table_maptrust_adjective_definitions} provides the classification and definition of the adjectives under their respective dimensions.

\begin{figure}[t]
\centering	
% only \linewidth for double col
\includegraphics[width=\linewidth ]{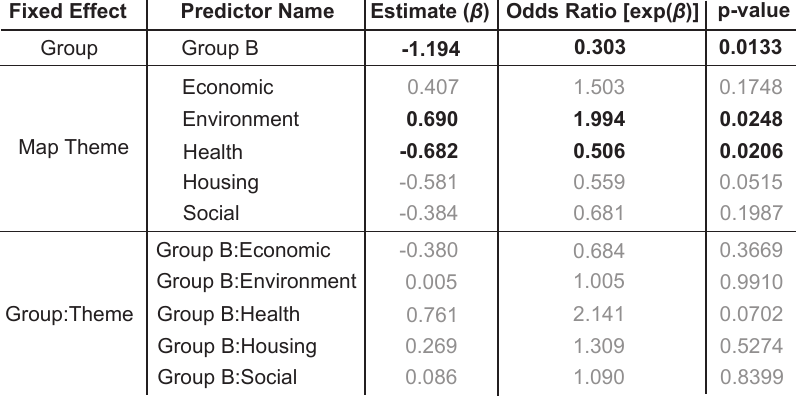}
\caption{The model includes three sets of predictors: Group (reference = Group A), Map Theme (reference = Crime), and their interaction terms (Group:Theme), which capture how the effect of each theme differs between groups. Estimates indicate the direction and magnitude of effects on the log-odds of assigning higher trust ratings, while Odds Ratios reflect the corresponding multiplicative change in odds. Significant effects (p < .05) are shown in bold.}
\label{fig:table_interaction_group_theme}
\Description{The table presents three sets of predictors and corresponding estimates, odds ratios, and p-values. The predictor sets include Group (with Group A as the reference), Map Theme (with the Crime theme as the reference), and their interaction terms.}
\end{figure}

\subsection{Exploratory Analysis: Map Theme Effects}
\label{appendixA3:map_theme_exploratory}

This supplementary analysis provides additional detail regarding exploratory patterns in how trust varied across map themes when uncertainty was visualized. As noted in Section~\ref{section6:limitations_future_work}, the study was not designed to test mechanisms underlying potential theme-level differences. Accordingly, the results reported here should be interpreted as descriptive patterns that may inform future research rather than as confirmatory evidence.

\begin{figure}[t]
\centering	
\includegraphics[width=0.65\linewidth ]{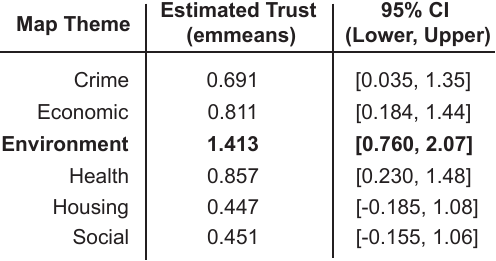}
\caption{Estimated marginal trust ratings for each map theme, computed using a model fit on Group B responses only (maps with uncertainty). The "Estimated Trust" column reports emmeans (estimated marginal means) for each theme. These values reflect the average predicted trust rating participants assigned to maps of each theme. The 95\% Confidence Interval (CI) column provides the lower and upper bounds. If the confidence interval includes zero (e.g., Housing), it suggests that the estimated trust rating is not statistically distinguishable from zero on the latent trust scale at the $\alpha$ = 0.05 level. The Environment theme elicited the highest estimated trust.}
\label{fig:table_emmeans_map_theme}
\Description{This table presents the estimated marginal trust ratings for each of the six map themes, calculated using responses from Group B participants who viewed maps with uncertainty visualization. The model used to generate these estimates controlled for the dominant uncertainty level in each map, isolating the effect of theme on trust. The first column lists the map themes evaluated. The second column, labeled "Estimated Trust," reports the estimated marginal mean (emmean) trust rating for each theme. These emmeans represent the average predicted value on the latent trust scale, not raw ratings, and provide insight into how each theme was perceived under uncertainty conditions. The third column provides the 95\% confidence interval for each estimate, indicating the range in which the true population value is expected to fall with 95\% certainty. If a confidence interval includes zero, such as in the case of the Housing theme, it implies that the estimated trust rating is not statistically distinguishable from zero on the latent scale at the 0.05 significance level. Among the themes, the Environment category stands out with the highest estimated trust rating and a confidence interval that excludes zero, indicating a robust positive effect. This analysis helps identify whether any specific themes maintain higher or lower trust ratings even when uncertainty is visualized.}
\end{figure}

To examine whether uncertainty visualization altered the extent to which trust differed by map theme, we fit a cumulative link mixed model (CLMM) that included Map Theme, Group, and their interaction as fixed effects, with Participant ID as a random intercept. Consistent with the main results, trust ratings in Group B (the uncertainty condition) were lower overall, and none of the interaction terms were statistically significant (Fig.~\ref{fig:table_interaction_group_theme}). This suggests that for Group B, uncertainty visualization had a relatively stable effect across themes.

To further contextualize the pattern observed in Group B, we computed estimated marginal means (emmeans)~\cite{searle1980population} using a CLMM fit on Group B responses only, with Map Theme and Dominant Uncertainty Level as fixed effects and Participant ID as a random intercept. These emmeans represent adjusted trust estimates for each theme, accounting for variation in exposure to uncertainty levels. As seen in Fig.~\ref{fig:table_emmeans_map_theme}, the Environment theme exhibited the highest estimated trust rating (EMMean = 1.413, 95\% CI [0.760, 2.07]).

Taken together, these exploratory findings suggest that uncertainty visualization may reduce variation in trust across map themes, with a possible but inconclusive indication that environmental topics elicited relatively higher trust under uncertainty. One speculative interpretation is that environmental data is more commonly presented with uncertainty information in public communication~\cite{joslyn2012uncertainty, ruginski2016non}, making viewers more familiar with or expectant of such representations. We encourage future work to more directly investigate the relationship between thematic context and uncertainty visualization using study designs specifically tailored to test such mechanisms. All model outputs, including pairwise contrasts, are provided in the Supplementary Materials for transparency.

\end{document}